\documentclass[a4paper,twocolumn,11pt,accepted=2023-11-23]{quantumarticle}
\pdfoutput=1

\usepackage[utf8]{inputenc}
\usepackage[english]{babel}
\usepackage{floatrow}

\newcommand{\sidecaption}[1]% #1 = label name
{\raisebox{\abovecaptionskip}{\begin{subfigure}[t]{1.6em}
  \caption[singlelinecheck=off]{}% do not center
  \label{#1}
\end{subfigure}}\ignorespaces}

\usepackage[export]{adjustbox}

\usepackage[T1]{fontenc}
\usepackage{amsmath}
\usepackage{hyperref}
\usepackage[normalem]{ulem}
\usepackage{tikz}
\usepackage{lipsum}

\usepackage{amsmath}
\usepackage{amssymb}
\usepackage{graphics}
\usepackage{stmaryrd}
\usepackage{comment}

\usepackage{ragged2e}
\DeclareCaptionJustification{justified}{\justifying}

\usepackage{subcaption} %
\usepackage[percent]{overpic}

\begin{document}

\newcommand{\reviews}[1]{#1}

\newcommand{\cO}{\mathcal{O}}
\newcommand{\cB}{\mathcal{B}}
\newcommand{\cC}{\mathcal{C}}
\newcommand{\RR}{\mathbb{R}}
\newcommand{\bK}{\mathbb{K}}
\newcommand{\cD}{\mathcal{D}}
\newcommand{\cU}{\mathcal{U}}
\newcommand{\cE}{\mathcal{E}}
\newcommand{\cc}{\mathfrak c}
\newcommand{\csu}{\mathfrak{su}}
\newcommand{\cM}{\mathcal{M}}
\newcommand{\cP}{\mathcal{P}}
\newcommand{\ba}{\hat{a}}
\newcommand{\bb}{\hat{b}}
\newcommand{\bc}{\hat{c}}
\newcommand{\bd}{\hat{d}}
\newcommand{\cG}{\mathcal G}
\newcommand{\cZ}{\mathcal Z}
\newcommand{\Cp}{\ket{\mathcal{C}_\alpha^+}}
\newcommand{\Cm}{\ket{\mathcal{C}_\alpha^-}}
\newcommand{\Cpd}{\bra{\mathcal{C}_\alpha^+}}
\newcommand{\Cmd}{\bra{\mathcal{C}_\alpha^-}}
\newcommand{\Cone}{\mathcal{C}_1}
\newcommand{\Ctwo}{\mathcal{C}_2}
\newcommand{\tph}{2\text{ph}}
\newcommand{\oph}{1\text{ph}}

\newcommand{\bra}[1]{\ensuremath{\langle#1|}}
\newcommand{\ket}[1]{\ensuremath{|#1\rangle}}
\newcommand{\ketbra}[2]{\ensuremath{|#1\rangle \langle #2|}}

\newcommand{\mytitle}{High-performance repetition cat code using fast noisy operations}

\title{\mytitle}

\author{Francois-Marie Le R\'egent}
\orcid{0000-0002-5229-7155}
\affiliation{Alice\&Bob, 53 boulevard du Général Martial Valin, 75015 Paris}
\affiliation{Laboratoire de Physique de l'Ecole Normale Supérieure, Ecole normale supérieure, MINES Paris, Université PSL, Sorbonne Université, CNRS, Inria, 75005 Paris}
\email{francois-marie.le-regent@inria.fr}

\author{Camille Berdou} 
\affiliation{Laboratoire de Physique de l'Ecole Normale Supérieure, Ecole normale supérieure, MINES Paris, Université PSL, Sorbonne Université, CNRS, Inria, 75005 Paris}

\author{Zaki Leghtas} 
\orcid{0000-0002-9172-1537}
\affiliation{Laboratoire de Physique de l'Ecole Normale Supérieure, Ecole normale supérieure, MINES Paris, Université PSL, Sorbonne Université, CNRS, Inria, 75005 Paris}

\author{J\'er\'emie Guillaud}
\orcid{0000-0001-6507-9344}
\affiliation{Alice\&Bob, 53 boulevard du Général Martial Valin, 75015 Paris}

\author{Mazyar Mirrahimi} 
\orcid{0000-0001-9471-6031}
\affiliation{Laboratoire de Physique de l'Ecole Normale Supérieure, Ecole normale supérieure, MINES Paris, Université PSL, Sorbonne Université, CNRS, Inria, 75005 Paris}

\maketitle

\begin{abstract} 
  Bosonic cat qubits stabilized by two-photon driven dissipation benefit from exponential suppression of bit-flip errors and an extensive set of gates preserving this protection. These properties make them promising  building blocks of a hardware-efficient and fault-tolerant quantum processor. In this paper, we propose a performance optimization of the repetition cat code architecture using fast but noisy CNOT gates for stabilizer measurements. This optimization leads to high thresholds for the physical figure of merit, given as the ratio between intrinsic single-photon loss rate of the bosonic mode and the engineered two-photon loss rate, as well as an improved scaling below threshold of the required overhead, to reach an expected level of logical error rate. Relying on the specific error models for cat qubit operations, this optimization exploits fast parity measurements, using accelerated low-fidelity CNOT gates, combined with fast ancilla parity-check qubits.  The significant enhancement in the performance is explained by: 1- the highly asymmetric error model of cat qubit CNOT gates with a major component on control (ancilla) qubits, and 2- the robustness of the repetition cat code error correction performance in presence of the leakage induced by fast operations. In order to demonstrate these performances, we develop a method to sample the repetition code under circuit-level noise that also takes into account cat qubit state leakage.
\end{abstract} 

\section{Introduction}

Bosonic encoding of quantum information is expected to lower the number of physical components required to perform quantum computations at scale~\cite{Joshi2021, Cai2021}. The crux of bosonic architectures is to leverage the infinite dimensional Hilbert space of a quantum harmonic oscillator (QHO) to implement some redundancy required for quantum error correction in a single physical component, an approach that has been coined ``hardware-efficient''~\cite{Mirrahimi2014}. Although these architectures are theoretically promising, operating such a concatenated ``bosonic code + discrete variable (DV) code'' below the threshold of the DV code is still experimentally challenging for current state-of-the-art superconducting platforms, thereby motivating subsequent research to improve the theoretical performance of these proposals.

In this work, we focus on the ``cat qubit + repetition code'' architecture~\cite{Guillaud_2019, Chamberland2022} with the objective of optimizing its error correcting capability. In this approach, the state of a QHO is confined through an engineered two-photon driven dissipative process to a two-dimensional subspace spanned by two coherent states $\ket{\pm\alpha}$, or equivalently by the
coherent superpositions of those two states, the Schr\"odinger cat states:
\begin{equation} \ket{\mathcal{C}_\alpha^\pm} := \mathcal{N}_\pm (\ket{\alpha}
\pm \ket{-\alpha}) \label{eq:encoding} \end{equation}
where $\mathcal{N}_\pm = (2(1\pm \exp(-2 |\alpha|^2)))^{-1/2} $ are normalizing constants. The computational states of this so-called cat qubit are given by
\begin{align*} \ket{0}_C &= (\ket{\mathcal{C}_\alpha^+} +
 \ket{\mathcal{C}_\alpha^-})/\sqrt{2} = \ket{\alpha} +
 \mathcal{O}(e^{-2|\alpha|^2}) \\ \ket{1}_C &= (\ket{\mathcal{C}_\alpha^+} -
 \ket{\mathcal{C}_\alpha^-})/\sqrt{2} = \ket{-\alpha} +
\mathcal{O}(e^{-2|\alpha|^2}). \end{align*}
 The engineered two-photon driven dissipative process can be effectively modelled by a Lindblad term of the form $L_{\tph}=\sqrt{\kappa_2}(\hat a^2-\alpha^2)$ (we refer to~\cite{Mirrahimi2014} for further details on how such a process can be engineered in a superconducting platform and how it stabilizes the cat qubit manifold).  This engineered process will be in competition with other coherent and incoherent processes tending to leak the QHO out of the cat qubit manifold or causing a drift in the manifold. Among these processes, the dominant one is the undesired single-photon loss, modelled by a Lindblad term of the form $L_{\oph}=\sqrt{\kappa_1}\hat a$.

The cat qubit stabilized by the two-photon driven dissipative process benefits from an intrinsic protection against bit-flip errors where the rate of such errors is exponentially suppressed with the mean number of photons $|\alpha|^2$~\cite{Lescanne2020,Berdou2022}. Relying on this protection, and the fact that it is possible to perform an extensive set of quantum operations preserving such an error bias, two of us proposed in~\cite{Guillaud_2019} to concatenate such an encoding with a repetition code to conceive a fault-tolerant architecture with a universal set of logical gates.

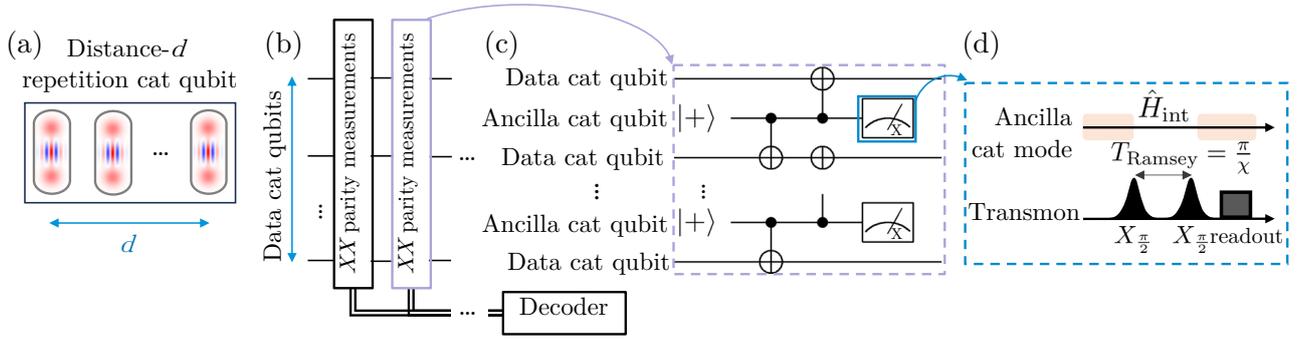
\begin{figure*}[t!]
  \centering
    \begin{overpic}[
      width=\textwidth,
      ]{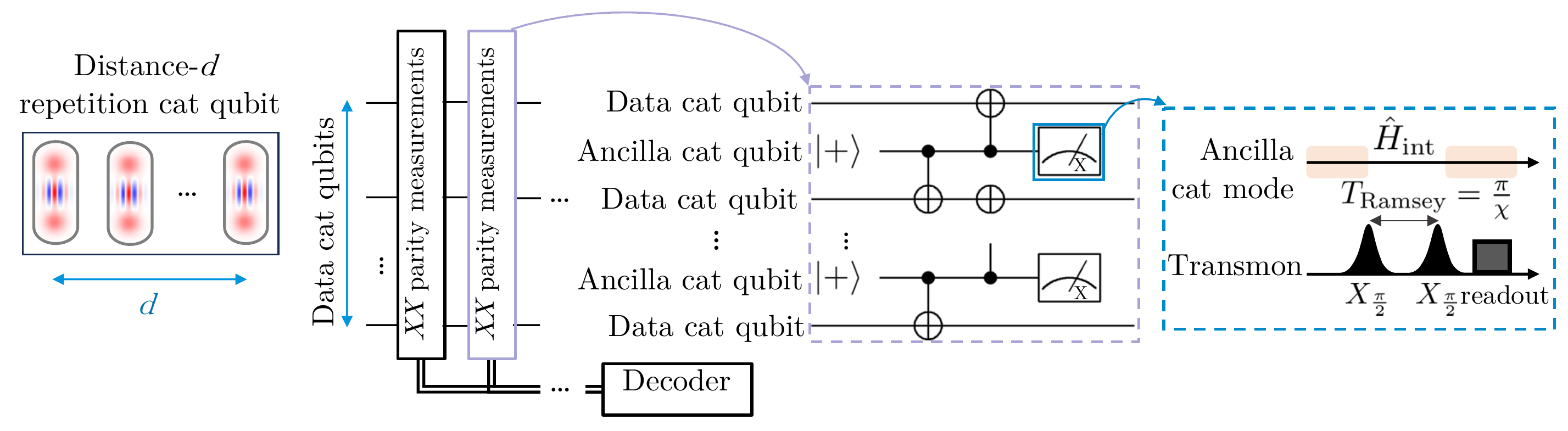}
    \put(0,23){(a)}
    \put(20,23){(b)}
    \put(37,23){(c)}
    \put(74,23){(d)}
    \end{overpic}
    \caption{
    \reviews{
    QEC circuit for the distance-$d$ repetition cat code.
    (a) a repetition cat code consists of a 1D array of stabilized cat qubits, where the phase-flip error correction is performed by repetitive $XX$ parity measurements between neighboring cat qubits.    
    (b) The QEC  on a distance-$d$ phase-flip repetition code is performed by repeated $XX$ parity measurements.
    (c)  Each QEC cycle is composed of $d-1$ $XX$ parity checks between neighboring cat qubits. 
    Each parity check is performed by applying bias-preserving CNOT gates between the two associated data cat qubits (as targets) and one ancilla cat qubit (as control). The ancilla cat qubit is finally measured in its $X$ basis.
    (d) The measurement of an ancilla cat qubit in the $X$ basis is performed through a photon number parity measurement. 
    This is done by turning off the two-photon dissipation mechanism on the ancilla cat qubit and following a Ramsey type experiment proposed in~\cite{Lutterbach1997} and realized in~\cite{Bertet2002, Sun2014}. 
    In this scheme the ancilla cat mode is coupled dispersively to a qubit, e.g. a transmon, where the interaction Hamiltonian is given by $\hat{H}_{\mathrm{int}}=\hbar \frac{\chi}{2} \hat{\sigma}_z \hat{a}^{\dagger} \hat{a}$. The Ramsey sequence consists then of applying two $\pi/2$-pulses on the qubit separated by a waiting time of $\pi/\chi$. During the waiting time, the qubit accumulates a phase of $0$ or $\pi$ depending on the photon number parity of the cat mode. The final readout of the qubit thus measures this observable.
    }
    \label{fig:repetition_cat_code}
    }
\end{figure*}

\reviews{We concatenate the cat qubits with a repetition cat code in order to protect it against phase flips.
The distance-$d$ repetition code encodes a single logical qubit into a 1D array of $d$ physical cat qubits, see~Fig.~\ref{fig:repetition_cat_code}(a).
To operate against phase flips, the code space is defined as the $+1$ common eigenspace of the $d - 1$ stabilizers of the set $ \mathcal{S} = \{ X_i X_{i+1}, i \in \llbracket 1~;~ d \rrbracket \} $
that are measured repeatedly~Fig.~\ref{fig:repetition_cat_code}(b).
The logical $|+\rangle_L$ and $|-\rangle_L$ states are given by tensor products of physical $\ket{\pm}$ states,  $| \pm\rangle_L:=$ $| \pm\rangle_C^{\otimes n}$ and the logical computational states are defined as:
$$
\begin{aligned}
|0\rangle_L & =\frac{1}{(\sqrt{2})^{n-1}} \sum_{j \in\{0,1\}^n,|j| \text { even }}|j\rangle_C \\
|1\rangle_L & =\frac{1}{(\sqrt{2})^{n-1}} \sum_{j \in\{0,1\}^n,|j| \text { odd }}|j\rangle_C.
\end{aligned}
$$
The quantum circuit to perform the stabilizer measurements is shown in Fig.~\ref{fig:repetition_cat_code}(c), where all the operations are implemented in a bias-preserving manner~\cite{Guillaud_2019}.
The $X$ logical operator of the ancilla cat qubit is then measured via a photon number parity measurement, displayed in Fig.~\ref{fig:repetition_cat_code}(d) (see also~\cite{Sun2014}).
}

The elementary figure of merit that quantifies the performance of this architecture is given by $\eta=\kappa_1/\kappa_2$, where $\kappa_1$ is the rate of undesired single-photon loss, and $\kappa_2$ corresponds to the rate of the engineered two-photon loss mechanism, stabilizing the cat qubit. More precisely, the performance of this architecture is quantified both in terms of a threshold for this figure of merit $\eta$ (called $\eta_{\text{th}}$ in this paper) below which the concatenation with the repetition code leads to an exponential suppression of phase-flip errors, and in terms of the physical resources required to operate the architecture at a given target error rate $\epsilon_{\mathrm{L}}$  for a fixed value of $\eta<\eta_{\text{th}}$.

In order to optimize the operation of this architecture, we investigate the acceleration of CNOT gates involved in stabilizer measurements. While the driven dissipative implementation of CNOT gate for  cat qubits, as detailed in~\cite{Guillaud_2019, Chamberland2022} and re-called in the next section, ensures the preservation of exponential bit-flip suppression, its acceleration can lead to significant increase in phase-flip error probability, decreasing the gate fidelity, and also to some  leakage out of the cat qubit subspace. Adding appropriate ``refreshing steps'' in the error correction logical circuit, countering the leakage of the cat qubit, we show that, despite  the degraded gate fidelity, the performance of error correction is significantly enhanced due to a faster measurement cycle. This enhancement is explained by two facts: first, the phase-flip error probability on the control (measurement) and target (data) cat qubits are highly asymmetric with the major contribution on the measurement qubits, and second, the residual leakage merely leads to short-range (both in time and space) measurement error correlations that affect  marginally the performance of error correction.  Through careful Monte Carlo simulations, with a circuit-level error model taking into account the impact of leakage, we show that operating the code in this fast gate regime achieves close-to-optimal performance, and that an application-wise relevant error probability of $\epsilon_{\mathrm{L}} = 10^{-10}$ per error correction cycle may be achieved with a repetition code of distance $25$ under a realistic hardware assumption $\eta = 10^{-3}$.

The paper is structured as follows. Section~\ref{sec:Motivation} summarizes the common principle exploited throughout this work. More specifically, we show how ``faster yet noisier'' gates are desirable in the context of error correction because of the high resilience of stabilizer codes to measurement errors (compared to errors damaging the encoded data); but lead in the context of dissipative cat qubits to important state leakage that needs to be carefully addressed. 
In Section~\ref{sec:fastgates}, we thoroughly analyze the impact of simple acceleration of CNOT gates on the error correction performance. In Section~\ref{sec:Asymmetry}, we further investigate the idea of asymmetry in the measurement and data cat qubit errors, by separating  the  time scales associated to their dynamics. More precisely, the idea that we exploit here is to use fast  measurement cat qubits with large single-photon and two-photon dissipation rates $\kappa_1$ and $\kappa_2$, and slow data cat qubits with smaller rates $\kappa_1$ and $\kappa_2$, but a similar ratio $\eta=\kappa_1/\kappa_2$.

\section{Repetition cat qubit, error model and leakage}
\label{sec:Motivation}

Similarly to the case of surface code with conventional qubits (e.g. transmons)~\cite{Fowler2012}, the error-correction performance of the ``cat qubit+repetition code'' architecture is mainly determined via the error probabilities of the CNOT gates involved in stabilizer measurements. In~\cite{Guillaud_2019}, inspired by~\cite{Puri2020}, two of us proposed an adiabatic implementation of the CNOT gate for the cat qubits stabilized by two-photon dissipation. While this implementation preserves the exponential suppression of bit-flip errors, the phase-flip errors can occur both due to intrinsic loss mechanism of the QHO and also due to higher order corrections to the adiabatic process. More precisely, on the one hand, the implementation needs to be slow enough with respect to the two-photon dissipation rate $\kappa_2$, so that the non-adiabatic effects do not induce significant phase-flip errors. On the other hand, this slow implementation leads to further phase-flip errors due to intrinsic single-photon loss of the QHO at rate $\kappa_1$. This leads to stringent requirements for the figure of merit $\eta=\kappa_1/\kappa_2$ to ensure a high-fidelity operation.

Throughout the past few years a certain number of proposals have targeted this issue~\cite{Xu2021,Putterman2022,Gautier2022,Ruiz2023,Xu2022Squeezed}. By various alterations of the dissipative process or the addition of the Hamiltonian confinements, these references aim at accelerating the operation of the bias-preserving gates, thus improving their fidelity. These modifications however usually come at the expense of more complex implementations, and sometimes also at the expense of losing the protection against bit-flip errors. As mentioned in the introduction, here we consider the possibility of using a low-fidelity operation by accelerating the operations for the original implementation~\cite{Guillaud_2019} and rather examine the impact of fast low-fidelity gates at the logical level~\cite{Xu2021}.  In this section, we start by   reminding the proposal of~\cite{Guillaud_2019} for realizing the CNOT gate and the associated error models.  We next discuss the expected performance of error correction at the repetition code level. Finally, we also  provide details on how the leakage induced by the finite gate time could limit this performance.

\subsection{CNOT gate and error model}\label{ssec:CNOTerrormodel}

In this paper, we denote the control cat qubit of the CNOT gate by the index $a$ (standing for the ancilla qubits for the repetition code stabilizer measurements) and the target one  by the index $d$ (standing for the data qubits). 
As proposed in~\cite{Guillaud_2019}, the CNOT gate can be implemented with a time-varying dissipative mechanism modelled by the master equation
\begin{equation}
  \frac{d\hat \rho}{dt}=\kappa_2\mathcal{D}[\hat L_a]\hat\rho+\kappa_2\mathcal{D}[\hat L_d(t)]\hat\rho-i[\hat H,\hat\rho].
\end{equation}
Here $\cD[\hat L](\hat\rho)=\hat{L} \hat{\rho}
\hat{L}^{\dagger}-\frac{1}{2}\left(\hat{L}^{\dagger} \hat{L}
\hat{\rho}+\hat{\rho} \hat{L}^{\dagger} \hat{L}\right)$, and 
$\hat L_a=\hat a^2-\alpha^2$ corresponds to regular two-photon driven dissipation for the ancilla mode $\hat a$ pinning its state to the manifold of cat states Span$\{\ket{\cC_\alpha^\pm}\}$, and 
$$
\hat L_d(t)=\hat{d}^{2}-\alpha^{2}+\frac{\alpha}{2}\left(e^{2 i \frac{\pi}{T}
t}-1\right)\left(\hat{a}-\alpha\right) 
$$ 
with $T$ the duration of CNOT gate, ensures a $\pi$-rotation of the target data mode $\hat d$ conditioned on the state of ancilla being in $\ket{-\alpha}$. The last term
$$
\hat{H} =\frac{\pi}{4 \alpha
T}\left(\hat a+\hat{a}^{\dagger}-2 \alpha\right)\left(\hat{d}^{\dagger}
\hat{d}-\alpha^{2}\right) 
$$
corresponds to a feed-forward Hamiltonian added to reduce the non-adiabatic errors induced by finite gate time.

The error model for such an implementation of the CNOT gate is detailed in~\cite{Guillaud_2019,Chamberland2022} and is briefly recalled here. This error model takes into account the non-adiabatic effects due to the finite gate time, as well as the errors induced by the undesired single photon decay of the ancilla and data modes. This undesired decay is modelled by the additional Lindbladian super-operators $\kappa_1\cD[\hat a]$ and $\kappa_1\cD[\hat d]$. As discussed in~\cite{Guillaud_2019, Chamberland2022}, the addition of other noise mechanisms such as thermal excitations or photon dephasing has little impact on these error models.  
While the bit-flip type errors remain exponentially suppressed as $\exp(-2|\alpha|^2)$, the probability of the phase-flip errors are given by
\begin{equation} p_{\mathrm{Z_a}} = |\alpha|^{2} \kappa_1 T + 0.159\frac{1}{
 |\alpha|^2 \kappa_{2} T} \\ \label{eq:error_model_symZ1} \end{equation}
 \begin{equation} p_{\mathrm{Z_d}} =p_{\mathrm{Z_a Z_d}} =\frac{1}{2}
 |\alpha|^{2} \kappa_1 T. \label{eq:error_model_symZ1Z2} \end{equation}

On the one hand, the probability of ancilla phase-flip errors $p_{\mathrm{Z_a}}$ comprises two parts. The first term corresponds to the errors induced by single photon loss and is proportional to the mean photon number of the cat state $|\alpha|^2$ and the gate duration $T$, and the second term to the errors induced by non-adiabatic effects. As analyzed in~\cite{Chamberland2022}, the probability of these errors scales inversely with $|\alpha|^2$ and $T$. The proportionality coefficient $0.159$ is obtained via a numerical fit, close to the estimated analytical value of $\pi^2/64$ derived in~\cite{Chamberland2022}. On the other hand, the data phase-flip errors $\mathrm{Z_d}$, as well as simultaneous data and ancilla errors $\mathrm{Z_a Z_d}$, are only induced by the single photon loss and their probability is therefore simply proportional to $|\alpha|^2T$. 

The gate time $T$ that minimizes the total phase-flip error probability of the CNOT gate $p_{\mathrm{CNOT}} = 
p_{\mathrm{Z_a}} + p_{\mathrm{Z_d}} + p_{\mathrm{Z_aZ_d}} $ is $T^{\star}=0.282/|\alpha|^2\sqrt{\kappa_1\kappa_2}$, which corresponds to a CNOT error probability 
\begin{equation} p_{\mathrm{CNOT}}^{*}= 
1.13\sqrt{\frac{\kappa_1}{\kappa_2}}.
\label{eq:pCNOT_opti} \end{equation}

\subsection{Expected performance of error correction~\label{ssec:QEC_Perf}}

\begin{figure*}[t!] \centering  
 \sidecaption{fig:leakage_cnot_reconvergence}   \raisebox{-\height}{\includegraphics[width=0.30\textwidth]{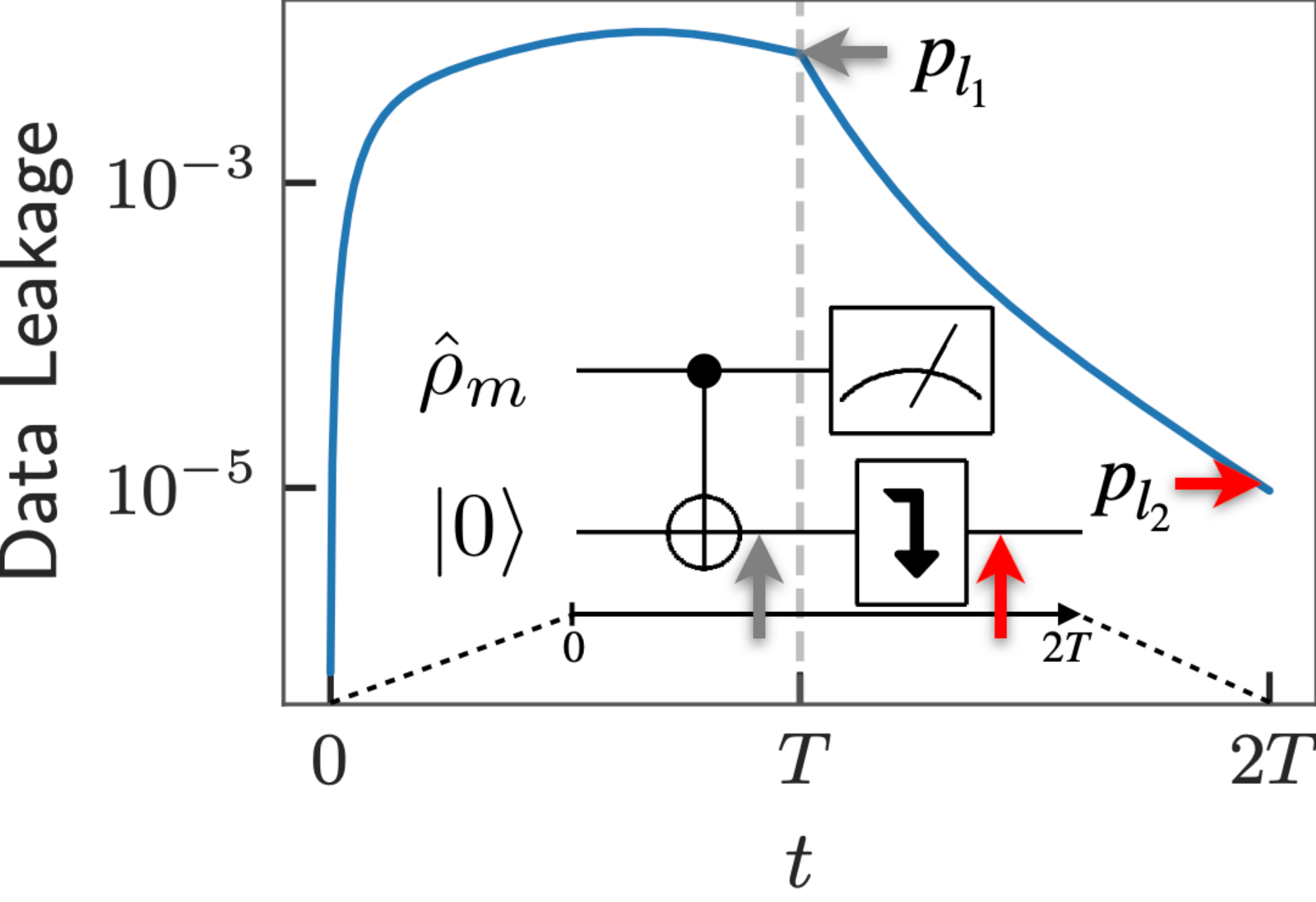}}
 \hfill
 \sidecaption{fig:leakage_vs_bitflip_changing_idle}   \raisebox{-\height}{\includegraphics[width=0.62\textwidth,]{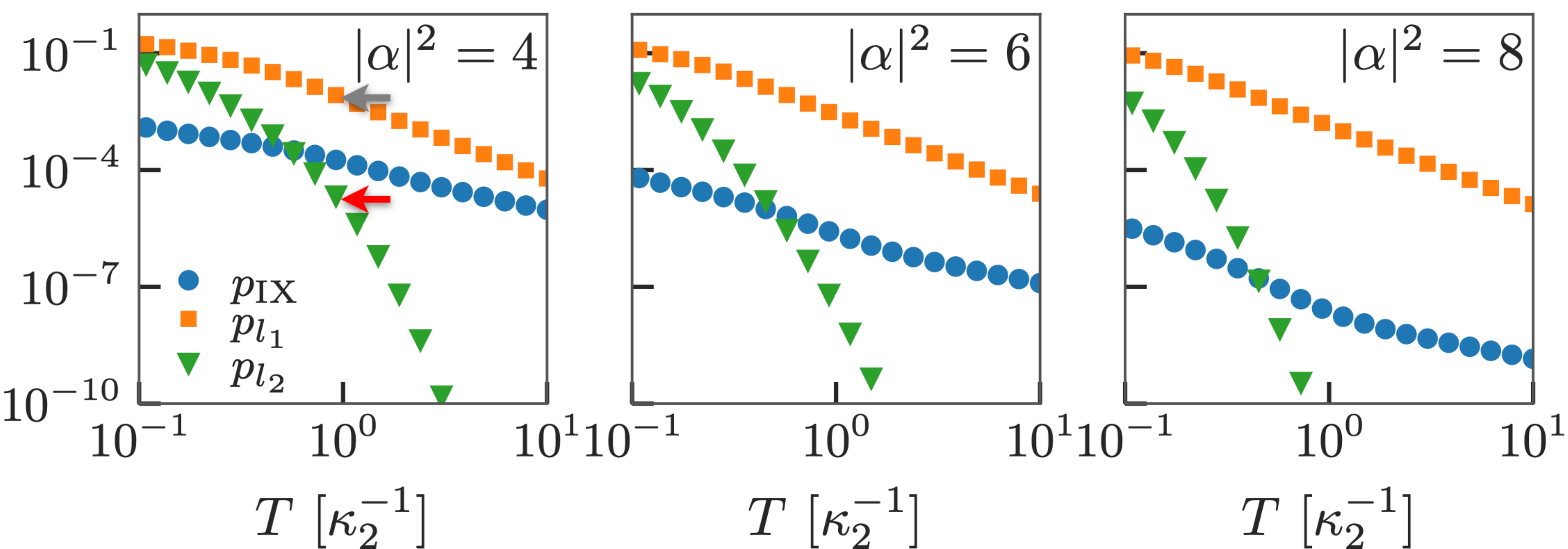}}
 \caption{
 (a) Leakage of the data qubit starting from the code space
  and during a CNOT  gate of duration $T=1/\kappa_2$ and a reconvergence step of the same duration for $|\alpha|^2=4$,
  $\eta = 10^{-3}$.
  \reviews{The
  inset displays both the quantum circuit used to mitigate the leakage via
  reconvergence and the locations of the leakage probabilities displayed in plot (b). More precisely, the gray arrow indicates that the leakage probability $p_{l_1}$ in plot (b) is given at this point in the circuit. The same indictaion holds for the red arrow and the leakage probability $p_{l_2}$.} 
  (b) Data leakage probabilities
  $p_{l_1}$ and $p_{l_2}$, before and after the refreshing step, and the bit-flip error probability $p_{\mathrm{IX}}$ on data qubit
  for different values of the gate duration $T$ (in units of $1/\kappa_2$) and mean photon population $|\alpha|^2$. The orange curves illustrate the leakage in the absence of the refreshing step and the green curves illustrate the leakage after a refreshing step of the same duration $T$ as the gate. Finally, the blue curves correspond to the bit-flip error probability. In the ``cat qubit+repetition code" architecture, we can safely neglect the leakage when its probability (green curve) is below the bit-flip probability (blue curve). \reviews{In these simulations, the ancilla used as the control qubit is initialized in 
  $\hat{\rho}_m = \frac{1}{2}( \ket{0}_C \bra{0}+ \ket{1}_C\bra{1}) =\frac{1}{2}( \ket{\mathcal{C}_\alpha^+} \bra{\mathcal{C}_\alpha^+}+ \ket{\mathcal{C}_\alpha^-}\bra{\mathcal{C}_\alpha^-})$}. This provides the average data leakage probability independently of the ancilla state.}  \end{figure*}

Here we study the CNOT gate from the perspective of its application in error syndrome measurements for phase-flip error correction. In this aim, the references~\cite{Guillaud_2021,Chamberland2022} perform Monte-Carlo simulations of the error correction logical circuit. These simulations are performed with a circuit-level error model in which all operations (gates, state preparations and measurements, and idling times) are noisy. 
More precisely, the CNOT gate errors are given by Eqs. \ref{eq:error_model_symZ1} and \ref{eq:error_model_symZ1Z2} with $T=T^{\star}$.
Furthermore, assuming that the ancilla preparation and measurement can be achieved in the same time $T^{\star}$, each ancilla preparation is accompanied by a phase-flip
error probability of $|\alpha|^2\kappa_1T^{\star}=0.282\sqrt{\kappa_1/\kappa_2}$ and
similarly, each ancilla measurement is faulty with probability 
$0.282\sqrt{\kappa_1/\kappa_2}$. Finally, the idle time during ancilla
measurement or preparation is accompanied by a phase-flip error probability of
$0.282\sqrt{\kappa_1/\kappa_2}$ in data qubits. The simulation results are
summarized in Figs.~\ref{fig:PRA_fitted} and~\ref{fig:PRA_overhead}. 

Denoting by $\eta := \kappa_1/\kappa_2$ the figure of merit for stabilized cat qubits, we roughly expect the scaling of the logical error
probability to be $p_{\mathrm{Z_L}}\propto
(\eta/\eta_{\text{th}})^{d/4}$, where $\eta_{\text{th}}$ refers to the fault-tolerance threshold and $d$ is the code distance. The power of $d/4$, instead of $d/2$, is explained by the fact that the physical error probabilities scales with $\sqrt{\eta}$. This expectation is confirmed by fitting the numerical results of Fig.~\ref{fig:PRA_fitted} to the ansatz
\begin{equation} p_{\mathrm{Z_{L}}}=ad\left(\frac{\eta}{\eta_{\text{th}}}\right)^{cd} \label{eq:fitZ_PRA} 
\end{equation} 
where we obtain the prefactor $a=7.7 \times 10^{-2}$, the exponential scaling $c=.258$ and the
phase-flip threshold $\eta_{\text{th}} =7.61 \times 10^{-3}$. These numerical results
are obtained by Monte Carlo simulations of the physical phase-flip errors and their
propagation in the circuit followed by a
minimum-weight perfect matching (MWPM) decoder~\cite{Fowler_PRL_2012, pymatching}. 

Here, we discuss these expected error correction performances in two limits. First, in the limit of $\eta\rightarrow \eta_{\text{th}}\approx 7.61\times 10^{-3}$, we note that the operation times $T^{\star}\rightarrow 3.23/|\alpha|^2\kappa_2$. This means that various gates are performed in times of order $1/\kappa_2$ or much shorter. As it will become clear in the next subsection, such short gate times lead to non-negligible leakage out of the code space that could lead to new challenges such as time-dependent and correlated error models. In order to overcome this problem, in the next subsection, we propose to add a qubit refreshment process acting as a leakage reduction unit (LRU). This however comes at the expense of a deterioration of the threshold as it increases the total duration of the QEC cycle. Next, we note that in the limit of $\eta\rightarrow 0$, the operation time $T^{\star}$, scaling as $1/|\alpha|^2\kappa_2\sqrt{\eta}$, becomes long with respect to the typical entropy evacuation time of $1/\kappa_2$. This is mainly to ensure a balanced reduction of error probability between data and ancilla qubits. In Section~\ref{sec:fastgates}, we argue that relaxing this requirement of balanced error probability reduction can lead to significantly better error correction performance. 

\subsection{Leakage and qubit refreshment step \label{ssec:LRU}}

The finite duration of the CNOT gates in the error correction circuit also leads to significant leakage out of the code space. Note that, contrary to the case of conventional qubits, and due to the continuous variable nature of encoding in cat qubits, the logical operations such as the CNOT gate perform rather well even in presence of leakage. The main issue is with a coherent build-up of the leakage leading to different error models for the operations in the logical circuit from one step to the other. More importantly, such a leakage could also lead to correlated errors in time and space  that could drastically limit the performance of the error correction. 

This leakage out of the code space is quantified by the mean value of the projector $\hat{\mathrm{P}}^{\perp} =
\hat{\mathrm{I}} - \hat{\mathrm{P}}$ with $\hat{\mathrm{P}} =
\ketbra{\mathcal{C}_\alpha^+}{\mathcal{C}_\alpha^+} +
\ketbra{\mathcal{C}_\alpha^-}{\mathcal{C}_\alpha^-}$ and $\hat{\mathrm{I}}$ is the identity. For instance, the optimal gate time $T^{\star}$ for $\eta=10^{-3}$ and $|\alpha|^2=8$ is close to $1/\kappa_2$. As it can be seen in the simulations of Fig.~\ref{fig:leakage_cnot_reconvergence}, this leads to a leakage out of the code space as large as $7.1 \times 10^{-3}$. 

A simple solution to handle leakage in the context of dissipative cat qubits consists in refocusing the cat qubit in the code space by letting it evolve under the action of two-photon driven dissipation. More precisely, each CNOT gate is followed by a qubit refreshing time during which the driven two-photon dissipation refocuses the leaked state to the code space. This simple process can be compared to more invasive LRUs considered for instance in the context of transmon qubits~\cite{Aliferis2007, Battistel2021, McEwen2021} that convert leakage into Pauli errors. One typical solution, implemented recently in an experiment~\cite{McEwen2021}, consists in adiabatically sweeping the qubit frequencies past a lossy resonator to swap excitations and go
back to the $(\ket{g}, \ket{e})$ manifold in every round of the QEC circuit ~\cite{Chen2021}. 

In Fig.~\ref{fig:leakage_vs_bitflip_changing_idle}, we compare the leakage rate after this qubit refreshing time with the bit-flip probability for different values of the cavity population $|\alpha|^2$. In these simulations, we consider a similar CNOT gate time and subsequent qubit refreshing time of $T$. By varying this duration $T$, we note that for $T\gtrsim 1/\kappa_2$, the leakage rate post refreshing time is below the bit-flip error probability and hence can be safely neglected. Therefore, we consider the duration $T=1/\kappa_2$ to be a lower bound
on the CNOT gate time one can use in the QEC circuit without introducing spurious effects due to leakage. This means that the logical circuit simulations of Fig.~\ref{fig:PRA_fitted} and the overhead estimation of Fig.~\ref{fig:PRA_overhead} from \cite{Guillaud_2021} need to be revised close to the threshold value for $\eta$, for which the gate duration $T^{\star}$ becomes too short.
Note that this lower bound could be reduced as $|\alpha|^2$ increases but we choose a conservative approach independent of the photon number.

Looking again at Figs.~\ref{fig:PRA_fitted} and~\ref{fig:PRA_overhead}, we need to consider the impact of adding these refreshing steps, therefore dealing with longer correction cycles. More precisely, we need to add a refreshing step of length $1/\kappa_2$ between the two CNOTs in one error correction cycle to avoid propagation and creation of correlated errors. Monte Carlo simulations indicate marginal change  in the overhead estimates with respect to Fig.~\ref{fig:PRA_overhead} for $\eta<10^{-3}$. But for $\eta$ larger than $10^{-3}$ the expected overhead can increase significantly. This can also be understood by looking at the color map on the curves of Fig.~\ref{fig:PRA_overhead}. For $\eta<10^{-3}$ the gate duration is larger than $1/\kappa_2$ which allows neglect the additional refreshing step.

\section{Accelerating QEC cycle with fast CNOTs} \label{sec:fastgates}

In the previous section, we noted that in the limit of $\eta\rightarrow 0$, a
CNOT operation time $T^{\star}$ ensuring its optimal fidelity becomes very long
compared to the entropy evacuation time of $1/\kappa_2$. We also argued that
this long operation time mainly secures a balanced reduction of error
probability between data and ancilla qubits. The idea that we pursue in this section
is that QEC is much more resilient to ancilla errors than data ones. More
precisely, the QEC tolerates finite measurement errors induced by ancilla phase-flips at the expense of a slightly degraded error threshold~\cite{Dennis2002}. This fact, further clarified in Subsection~\ref{ssec:pheno},
motivates the choice of accelerating the CNOT gates to a
minimal gate time of $1/\kappa_2$ equivalent to the time needed for leakage
removal. Relying on the asymmetric error model of the CNOT gate, discussed in
Subsection~\ref{ssec:CNOTerrormodel}, such a reduction of the gate time leads to a
reduced phase-flip error probability of data qubits due to single photon loss, at the
expense of increasing ancilla qubits phase-flip error probability induced by non-adiabatic
effects. In Subsection~\ref{ssec:overhead_reducedtime}, we show that this
faster cycle time drastically improves the error correction
performance scaling with $\eta$.

\subsection{Resilience of QEC to ancilla errors\label{ssec:pheno}} 

Following the discussion of Subsection~\ref{ssec:CNOTerrormodel}, with the choice of $T^{\star}$ as the CNOT gate time, the error probability for the ancilla and data qubits (neglecting the correlation for the simultaneous data and ancilla errors) are given by 
$$
p_{\mathrm{Z_a}}=0.987\sqrt{\eta},\qquad p_{\mathrm{Z_d}}=0.282\sqrt{\eta}.
$$
Taking into account the ancilla preparation and detection errors, as well as the idle time data errors, this gives rise to a phenomenological error model~\cite{Dennis2002} with data error probability per cycle given by $p=1.128\sqrt{\eta}$ and measurement error probability given by $q=2.538\sqrt{\eta}$. In particular, in the limit where $\eta\rightarrow 0$, both these error probabilities also tend to zero with $\sqrt{\eta}$. This explains the threshold curves provided in Fig.~\ref{fig:PRA_fitted} where $p_{\mathrm{Z_L}}$ scales with $\eta^{d/4}$ in the limit of small $\eta$. 

\begin{figure}
  \centering
    \begin{overpic}[
      width=0.95\textwidth,
      ]{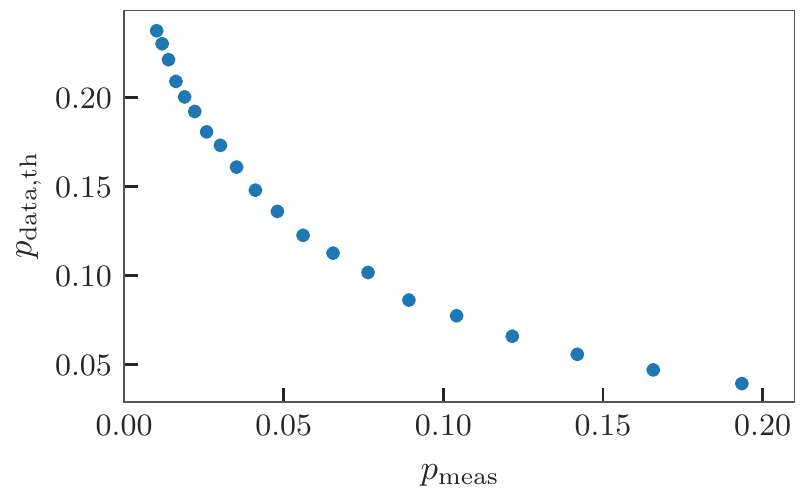}
    \put(55,28){\includegraphics[scale=.9]{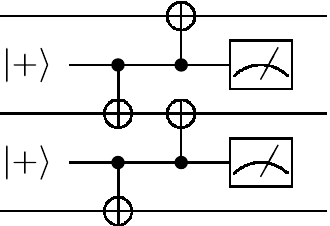}}
    \put(56, 27.7){\huge \color{red}{$\star$}}
    \put(56, 39.3){\huge \color{red}{$\star$}}
    \put(56, 51){\huge \color{red}{$\star$}}
    \put(83.5, 33.5){\huge \color{blue}{$\star$}}
    \put(83.5, 45){\huge \color{blue}{$\star$}}
    \put(54, 57.5){\color{red}{$p_{\mathrm{data}}$}}
    \put(80, 57.5){\color{blue}{$p_{\mathrm{meas}}$}}
    \end{overpic}
    \caption{
    \reviews{
    Error probability threshold $p_{\mathrm{data, th}}$ of the repetition code using the phenomenological error model when the measurement error probability is fixed to $p_{\mathrm{meas}}$.
    One stabilizer measurement cycle is shown in the inset where the locations of errors are highlighted: data errors with probability $p_{\mathrm{data}}$ (shown as red stars) and measurement errors with probability $p_{\mathrm{meas}}$ (shown as blue stars).
    The circuit is repeated $d$ times.
    The circuit is sampled $10^7$ times, or until $10^3$ logical errors have occurred, whichever comes first.
    The error bars are smaller than the data point sizes.
    }
    }
    \label{fig:Pheno_threshold_curve}
\end{figure}
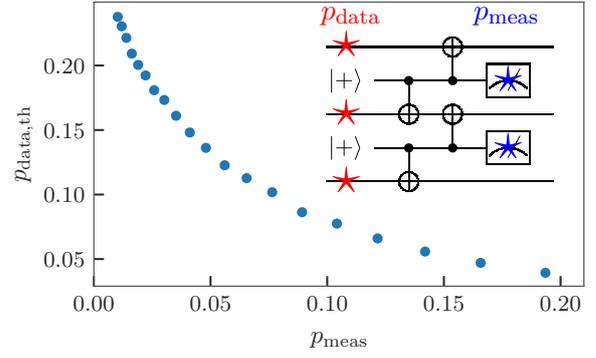

As explained earlier, the idea that we pursue in this section is to reduce the operation times to $T=1/\kappa_2$ instead of $T^{\star}$. Following once again the discussion of Subsection~\ref{ssec:CNOTerrormodel}, the ancilla and data qubit error probabilities are now given by
$$
p_{\mathrm{Z_a}}=0.159\frac{1}{|\alpha|^2}+1.5|\alpha|^2\eta,\qquad p_{\mathrm{Z_d}}=|\alpha|^2\eta. 
$$
We furthermore assume that the ancilla preparation and measurement, as well as data idle time error probabilities are given by $|\alpha|^2\kappa_1 T=|\alpha|^2\eta$. In order to avoid leakage induced errors, we further consider a qubit refreshing time (LRU) of $1/\kappa_2$ between two CNOT operations in one cycle (see the inset of Fig.~\ref{fig:QEC_sym_nbar8_fitted}). During this qubit refreshing time, we need to consider an additional error probability of $|\alpha|^2\eta$ for both ancilla and data qubits. Now, for any value of the mean photon number $|\alpha|^2$, as $\eta$ goes to zero, the data error probability $p=5|\alpha|^2\eta$ tends to zero proportionally to $\eta$ (to be compared to $\sqrt{\eta}$ in the previous case), but the ancilla error probability $q=0.318/|\alpha|^2+6|\alpha|^2\eta$ converges to a fixed non-zero value given by $0.318/|\alpha|^2$. 
As discussed in~\cite{Dennis2002}, for such a phenomenological error model where the measurement error probability is fixed, it is still possible to find a threshold for the data errors. 
In Fig~\ref{fig:Pheno_threshold_curve}, we plot the threshold $p_{\text{data, th}}$ for data error probability as a function of a fixed measurement error probability $p_{\text{meas}}$. 
More precisely, for fixed values of $p_{\text{meas}}$ between 1\% and $20\%$, we numerically calculate $p_{\text{data, th}}$ such that the logical error probability $p_{\mathrm{Z_L}}$ after a MWPM decoding scales as $ p_{Z_{\mathrm{L}}} = a d (p_{\text{data}}/p_{\text{data, th}})^{c(d+1)}$, with $c\approx 0.5$.
\reviews{\reviews{
See the Appendix for further details on the QEC circuit sampling and the fitting procedure to obtain the threshold values}.
}

We note that, with a significant measurement error probability of $10$ to $20\%$, we can still expect error thresholds of a few percents on data qubits. Furthermore, increasing the measurement error in this range, we only slightly decrease the data error threshold value. This threshold is thus quite resilient to measurement errors. 
\reviews{
Also note that for $p_{\mathrm{meas}}=10\%$, the data threshold $p_{\mathrm{data}}$ is smaller than the value (10\%) obtained in the context of the repetition code with symmetric phenomenological error model ($p_{\mathrm{meas}}=p_{\mathrm{data}}$)~\cite{Dennis2002}. This is explained by the fact that the $p_{\mathrm{meas}}$ is fixed to $10\%$ even in the asymptotic regime $p_{\mathrm{data}} \ll p_{\mathrm{data, th}}$ where the threshold is computed.
}

\subsection{Logical circuit simulations and overhead estimates \label{ssec:overhead_reducedtime}}

Similarly to the Subsection~\ref{ssec:QEC_Perf}, here we numerically simulate the error correction circuit plotted in the inset of Fig.~\ref{fig:QEC_sym_nbar8_fitted}. In this logical circuit, at each QEC cycle, the ancillas are prepared in the state $\ket{+}_C$ over a time duration of $1/\kappa_2$ and are also measured along their $X$ axis over a similar duration. The CNOT gates are also performed over the same duration of $1/\kappa_2$, and if not followed by an ancilla preparation and measurement step, we consider an additional qubit refreshing step of duration $1/\kappa_2$ to refocus the qubit state on the cat manifold, avoiding leakage-induced problems. Similarly to the previous subsection, the phase-flip error probabilities for the qubit preparation, measurement and idling steps are given by $p_{\mathrm{Z_a}}=p_{\mathrm{Z_d}}=|\alpha|^2\eta$. Also the phase-flip error probabilities for the CNOT gates are given in~\eqref{eq:error_model_symZ1} and~\eqref{eq:error_model_symZ1Z2} with $T=1/\kappa_2$. 

The results of these simulations for the particular choice of mean photon number $|\alpha|^2=8$ and several code distances $d$ are plotted in Fig.~\ref{fig:QEC_sym_nbar8_fitted}. These results are to be compared to the case of the choice of $T^{\star}$ as the duration of the operations, plotted in Fig.~\ref{fig:PRA_fitted} and discussed in Subsection~\ref{ssec:QEC_Perf}. We can see that for this value of $|\alpha|^2=8$, the threshold $\eta_{\text{th}}$ has decreased from $7.6 \times 10^{-3}$ to $2.3 \times 10^{-3}$. This can be simply explained with the arguments in the end of Subsection~\ref{ssec:LRU}. Indeed, this threshold is over-estimated for the case $T=T^{\star}$ as close to this choice, the CNOT gates could induce important leakage that is neglected in the simulations of Fig.~\ref{fig:PRA_fitted}. However, the important observation is that below the new threshold value $\eta_{\text{th}}\approx 2.3 \times 10^{-3}$, in the regime where we benefit from the exponential suppression of the logical phase-flip with the code distance, the coefficient in the exponent has nearly doubled going from $c=2.58 \times 10^{-1}$ (close to one quarter) to $c=4.4\times 10^{-1}$ (close to one half).

So far, we have only considered the logical phase-flip error. In order to estimate the required overhead to reach a certain logical error rate, the bit-flip errors need to be taken into account. In the logical circuit (inset of Fig.~\ref{fig:QEC_sym_nbar8_fitted}), all operations have non zero bit-flip error probability. Nevertheless, by far, the most significant contribution to the bit-flip error is due to the CNOT gates and therefore we neglect the contribution of the other operations. For the CNOT gate, with the parameter $\eta$ in the typical range of values between $10^{-5}$ to $10^{-2}$ considered here, the probability of bit-flip type errors numerically fits the following ansatz $p_{\mathrm{X}}^{\mathrm{CNOT}}= 0.5 \times e^{-2 |\alpha|^2}$.
Here $p_{\mathrm{X}}^{\mathrm{CNOT}}$ sums over all the possible error mechanisms leading to a bit-flip: $X_1$, $X_2$, $Y_1$, $Y_1X2$, $X_1X_2$, $Z_1X_2$, $Y_2$, $X_1Y_2$,
$X_1Z_2$, $Y_1Y_2$, $Y_1Z_2$ and $Z_1Y_2$. Also, while the first term corresponds to bit-flip errors induced by single photon loss at rate $\kappa_1$, the second term (independent of $\eta$) corresponds to non-adiabatic bit-flip errors. Looking at the QEC circuit plotted in the inset of Fig.~\ref{fig:QEC_sym_nbar8_fitted}, an upper bound for the total bit-flip error probability per QEC cycle is therefore given by $p_{\mathrm{X_L}} = 2(d-1)p_{\mathrm{X}}^{\mathrm{CNOT}}$. 

\begin{figure*}[t!] \centering
 \sidecaption{fig:PRA_fitted}   \raisebox{-\height}{\includegraphics[width=0.46\textwidth]{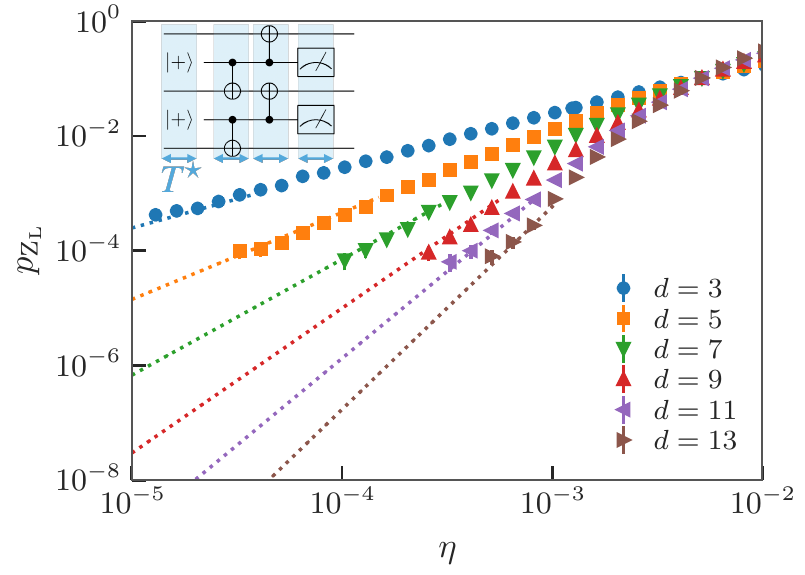}}
  \sidecaption{fig:QEC_sym_nbar8_fitted}   \raisebox{-\height}{\includegraphics[width=0.46\textwidth]{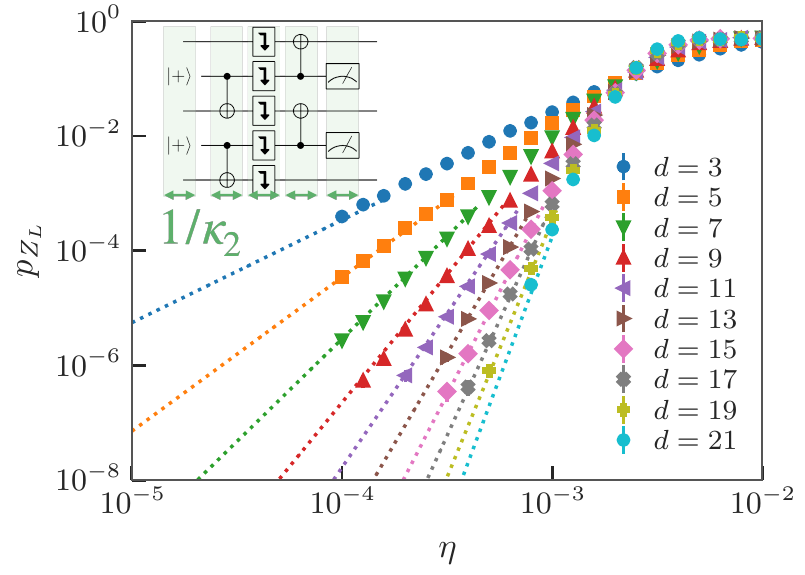}}
\caption{
  Probability that the error correction based on the parity measurement circuit displayed in
 the inset induces a logical $Z_L$
 error on the repetition cat qubit after the correction is performed. The
 dotted lines correspond to the asymptotic regime and fit the empirical
 scaling formula $p_{\mathrm{Z_{L}}}=ad\left(\frac{\eta}{\eta_{\mathrm{th}}}\right)^{cd}$\reviews{, see App.\ref{sec:mc}}. (a) These simulations correspond to the case where the operation times that are fixed to $T^*$ and neglect the leakage induced by fast operations (occurring for $\eta\gtrsim 10^{-3}$). For the scaling formula, we find  $a=7.7 \times 10^{-2}$, $c=.258$ and $\eta_{\mathrm{th}}=7.6 \times 10^{-3}$. (b) These simulations correspond to the case where the operation times are fixed to $1/\kappa_2$ and where we add refreshing steps between CNOT gates to remove the leakage, and $|\alpha|^2 = 8$. In this case, we find $a=3.2\times 10^{-2}$, $c=.44$ and $\eta_{\mathrm{th}}=2.3\times 10^{-3}$.
 \label{fig:fits}
 }
  \end{figure*}
 
Now, the overall logical error probability per error correction cycle $\epsilon_{\mathrm{L}}$ can be upper bounded by $p_{\mathrm{X_L}}+p_{\mathrm{Z_L}}$. In Fig.~\ref{fig:overhead_fixedCNOT}, we plot, as a function of $\eta$, the minimum values of the code distance $d$ and of the mean photon number $|\alpha|^2$ leading to a logical error rate $\epsilon_{\mathrm{L}}=10^{-5}$, $10^{-7}$ and $10^{-10}$. This is to be compared to the choice of $T=T^{\star}$ for the operations in the logical circuit, plotted in Fig.~\ref{fig:PRA_overhead}. Once again one should note that the estimated overhead in this latter case needs to be taken with precaution as for larger values of $\eta \gtrsim 10^{-3}$, strong leakage has been neglected in these simulations. The addition of a qubit refreshing time as in Subsection~\ref{ssec:LRU}, would bring the required overhead closer to what is estimated in Fig.~\ref{fig:overhead_fixedCNOT} for these values of $\eta \gtrsim 10^{-3}$. More interestingly, for smaller values of $\eta$ below $10^{-4}$, the required code distance is smaller because of the better scaling of $p_{\mathrm{Z_L}}$ with $\eta$, as explained above. For example, for $\eta = 10^{-5}$, a logical error rate of $\epsilon_{\mathrm{L}}=10^{-10}$ is achieved using a repetition of $9$ cat qubits of size $|\alpha|^2\approx14$ versus a repetition code distance of 15 cat qubits of size $|\alpha|^2\approx13$ with the prior choice $T=T^{\star}$.
The higher photon number can be explained by a higher bit-flip error probability coming from non adiabaticity in the context of faster gates, as can be seen in the insets of Figs. \ref{fig:PRA_overhead} and \ref{fig:overhead_fixedCNOT}.

\begin{figure*}[t!] \centering   
\sidecaption{fig:PRA_overhead}   \raisebox{-\height}{\includegraphics[width=0.48\textwidth]{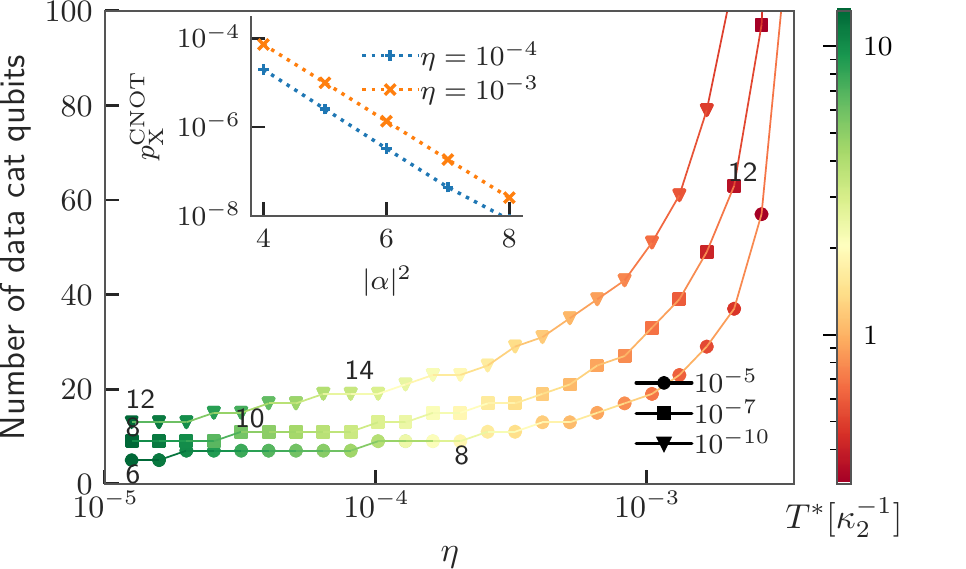}}
\sidecaption{fig:overhead_fixedCNOT}   \raisebox{-\height}{\includegraphics[width=0.42\textwidth]{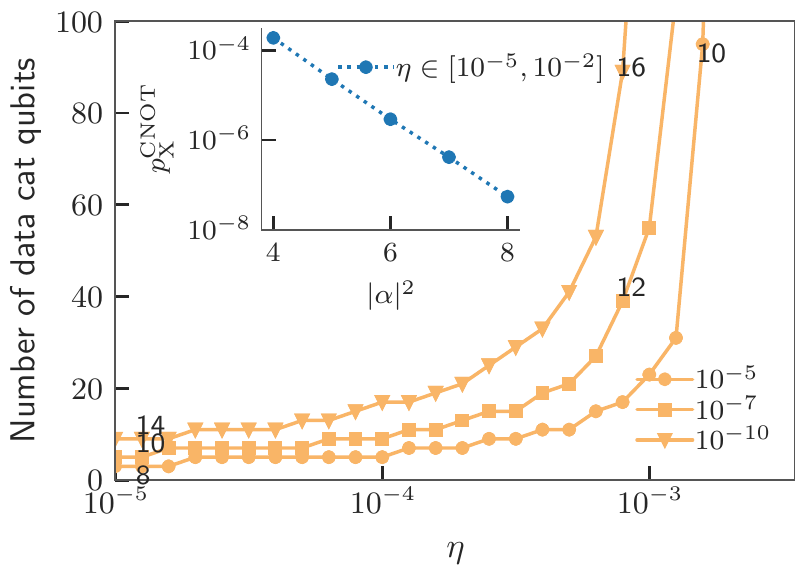}}
\caption{
  Estimated number of cat qubits per repetition cat qubit used as a
 quantum memory, versus the physical figure of merit $\eta=\kappa_1/\kappa_2$. The different curves correspond to different values of
 $\epsilon_L$ the target
 logical error probability per QEC cycle, and the numbers on the curves give an indication of  the required mean number of photons $|\alpha|^2$.
 \reviews{
 These points are found through a fitting procedure to extrapolate per-cycle phase-flip error rates with increasing code distance $d$.}
  (a) This plot corresponds to the case of Fig.\ref{fig:PRA_fitted} where the operation times are fixed to $T^*$ and the leakage is neglected. The color plot shows the value of operation time $T^*$ which changes here with $\eta$ and $|\alpha|^2$. The overhead is underestimated for $\eta\gtrsim 10^{-3}$ as the operation time $T^*$ becomes short in front of $1/\kappa_2$, thus leading to an important leakage out of the cat qubit subspace. The inset shows the total bit-flip error probability versus mean number of photons for two typical values of $\eta$ during a CNOT gate. Here, the dotted lines correspond to the numerical fit $p_X^{\mathrm{CNOT}}=\left(5.58 \sqrt{\eta}+1.68 \eta\right) e^{-2 |\alpha|^2}$~\cite{Chamberland2022}.
 (b) This plot corresponds to the case of Fig.\ref{fig:QEC_sym_nbar8_fitted}, where the operation times are fixed to $1/\kappa_2$ and extra refreshing steps are added to suppress leakage.   The inset shows the total bit-flip error probability versus mean number of photons with a rate ranging from $\eta = 10^{-2}$ to $\eta =10^{-5}$ during a CNOT gate. Here, the dotted line corresponds to the numerical fit $p_{\mathrm{X}}^{\mathrm{CNOT}}= 0.5 \times e^{-2 |\alpha|^2}$.
 \label{fig:overhead}
 } 
\end{figure*}

\section{Accelerating measurement cycle with fast ancilla qubits}
\label{sec:Asymmetry}

In the previous section, we showed that the error correction performance of the ``cat qubit+repetition code'' architecture could be improved by accelerating the CNOT gates. In this section, we explore a different idea to further accelerate the error syndrome measurements. 

The key fact we exploit here is that, from an experimental point of view, the difficult quantity to minimize is the ratio between the undesired single-photon loss (or other decoherence channels of the harmonic oscillator) and the engineered two-photon dissipation rate $\eta = \kappa_1 / \kappa_2$, while there is some flexibility to set the absolute value of these quantities ($\kappa_1$ and $\kappa_2$) by varying the specific circuit design of the cat qubit. The hint behind this fact is that circuit designs leading to high-Q modes with very low $\kappa_1$ usually rely on a strict isolation of the mode, making it harder to get a strong non-linear coupling to this mode. As a consequence, it is harder to engineer a two-photon exchange between this high-Q mode and a low-Q buffer mode, ultimately limiting the strength of the engineered two-photon dissipation $\kappa_2$. Motivated by this observation, we now assume an asymmetry between the dissipative rates of the ancilla and data cat qubits $\Theta:=\kappa_2^a/\kappa_2^d>1$, while keeping the value of $\eta=\kappa_1^a/\kappa_2^a=\kappa_1^d/\kappa_2^d$ fixed. 

This section is structured as follows. First, we demonstrate how this extra freedom in the system parameters can be exploited to obtain a drastic improvement of error correction performance. The regime achieving this performance requires implementing fast CNOT gates with respect to the data cat qubit stabilization time $1/\kappa_2^d$, resulting in an important state leakage for the data qubits. Next, we investigate the two spurious effects induced by this leakage: the bit-flip errors induced by leakage accumulation and correlated measurement errors for phase-flip error correction. Via thorough numerical simulations including these effects, we argue that they are not detrimental to the operation of the code in this regime. In Subsection~\ref{ssec:asym_correlation}, by using an appropriate basis for the Hilbert space of the data cat qubits, we show that one can  use a classical model for measurement error correlations. This observation makes the full Monte Carlo simulations of the logical error correction circuit including measurement correlations numerically tractable. Building on such numerical simulations, in Subsection~\ref{ssec:overhead_asym}, we compute the error correction overheads for the implementation presented in this section.

\begin{figure} \centering    
  \includegraphics[width=0.95\textwidth]{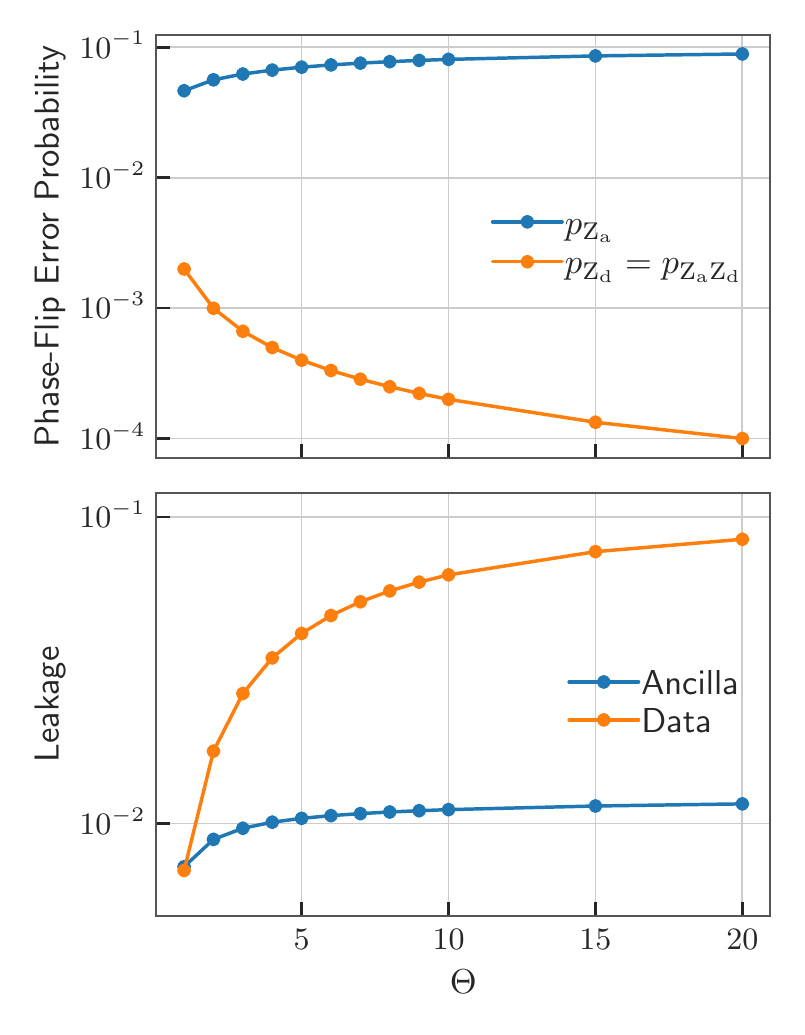}\label{fig:leakage_vs_asym}
 \caption{Ancilla and data phase-flip error probability (top) and of the leakage (bottom) as a function of the asymmetry $\Theta = \kappa_2^a / \kappa_2^d$ for a
cavity population of $|\alpha|^2 = 4$ and figure of merit $\eta = 10^{-3}$.} \label{fig:CNOT_em_vs_asym}
\end{figure}

\subsection{Asymmetric phase-flip error model and state leakage\label{ssec:asym}}

As detailed in subsection~\ref{ssec:CNOTerrormodel}, performing the CNOT gate in finite time creates non-adiabatic phase-flip errors on the ancilla qubit with probability $0.159 / |\alpha|^2 \kappa_{2} T$. When the ancilla and data cat qubits have different stabilization rates, $\Theta > 1$, one may wonder whether the gate should be adiabatic with respect to the slowest of the two timescales $1/\kappa_2^d$, or if it suffices to be adiabatic with respect to the fast timescale $1/\kappa_2^a$. We check the latter by numerically simulating the following evolution implementing a CNOT gate (in presence of single-photon loss)
 \begin{equation} \label{eq:ME_init} \begin{split} \frac{d \hat{\rho}}{d t}\ =\ &
  \kappa_{2}^a
  \mathcal{D}[\hat{a}^{2}-|\alpha|^2]\hat{\rho}+\kappa_2^d
  \mathcal{D}[\hat{L}_d(t)] \hat{\rho} \\
  &+\kappa_1^a\mathcal{D}[\hat{a}]\hat{\rho}+\kappa_1^d
 \mathcal{D}[\hat{d}] \hat{\rho}-i[\hat{H},
\hat{\rho}] \\ \end{split} \end{equation}
As discussed previously, we assume $\kappa_1^a/\kappa_2^a=\kappa_1^d/\kappa_2^d=\eta$, and we vary the asymmetry value $\Theta=\kappa_2^a/\kappa_2^d$, and the gate time is set to $T_{\mathrm{CNOT}}=1/\kappa_2^a$. The resulting phase-flip error probabilities are shown in Fig.~\ref{fig:CNOT_em_vs_asym}. The non-adiabatic phase-flip errors on the ancilla qubit only slightly increase with the asymmetry $\Theta$, which indicates that the gate time $T_{\mathrm{CNOT}}=1/\kappa_2^a$ is sufficiently slow with respect to the timescale of the ancilla qubit. On the data cat qubit, however, the phase-flips are only caused by single photon loss, which scale with the CNOT gate time as $p_{\mathrm{Z_d}} = |\alpha|^2 \kappa_1^d T_{\mathrm{CNOT}}/2 = |\alpha|^2 \eta / 2\Theta$. Therefore, for a fixed value of $\eta$ and for a gate time $ T_{\mathrm{CNOT}}=1/\kappa_2^a$, increasing the asymmetry results in decreasing the data phase-flip error probability. 

Given the considerations of Subsection~\ref{ssec:pheno} on the resilience of QEC to measurement errors, it is expected that increasing the ratio $\Theta$ between the ancilla and data stabilization rates leads to an improvement of the error correction performance, as it results in a linear reduction of data phase-flip errors at the cost of a slight increase in measurement errors. At a heuristic level, this can be understood from the fact that the performance of QEC depends on the typical time at which syndrome information is extracted (here, $\propto 1/\kappa_2^a$) versus the typical time at which errors occur ($\propto 1/\kappa_1^d$)
$$
\frac{\text{QEC cycle time}}{\text{quantum coherence time}} \propto \frac{\kappa_1^d}{\kappa_2^a} = \frac{\eta}{\Theta}.
$$
Thus, by leveraging the system asymmetry $\Theta$, one can hope to achieve a higher threshold value for $\eta$.

While this approach seems promising, it creates an important issue that needs to be addressed. Indeed, by fixing the gate time according to the stabilization rate of the fast ancilla qubits, the data cat qubits see a much faster dynamics than their confinement rate, $T_{\mathrm{CNOT}} = 1 / \Theta\kappa_2^d  \ll 1/\kappa_2^d$. As argued in Subsection~\ref{ssec:LRU}, such fast gates lead to an important amount of leakage outside the code space. This is numerically investigated in Fig.~\ref{fig:CNOT_em_vs_asym}, and as expected, increasing the system asymmetry, while fixing the gate time to be the inverse of the ancilla qubit stabilization rate, results in a constant leakage on the ancilla qubit but leads to an important amount of leakage on the data qubit. 

As previously discussed in Subsection~\ref{ssec:LRU}, this can lead to two problems:  leakage induced bit-flips on data qubits, and correlations in the measurement errors compromising the functioning of the repetition code error correction. In the next subsection, we investigate these two effects. 

\subsection{Numerical investigation of leakage-induced bit-flips and measurement error correlations\label{ssec:num_asym}}

In order to investigate the effect of data leakage as a function of increased system asymmetry, we perform numerical simulations of repeated logical $X$ measurements according to the circuit depicted in Fig.~\ref{fig:z_correlations_circuit}. In this simulation, we focus on the non-adiabatic effects by neglecting other noise sources (i.e. we assume $\kappa_1^a = \kappa_1^d = 0$). The measurement is repeated $\Theta$ times (using integer values of the asymmetry $\Theta$ for simplicity) as the system asymmetry is increased, such that the total simulated time $T = 1/\kappa_2^d$ is fixed.

To check the impact of leakage on the bit-flip errors, the data cat qubit is initialized in $|0\rangle_C=(\ket{\cC_\alpha^+}+\ket{\cC_\alpha^-})/\sqrt{2}$ and the ancilla in the mixed state $\rho_a = (|+\rangle_C\langle+| + |-\rangle_C\langle-|)/2=(\ket{\cC_\alpha^+}\bra{\cC_\alpha^+}+\ket{\cC_\alpha^-}\bra{\cC_\alpha^-})/2$. The particular choice of fully mixed initial state for the ancilla qubit provides an average of the bit-flip error probabilities over all possible initial states. Also the dynamics for the data cat qubit is symmetric, which explains the choice of initial state $\ket{0}$. For each value of $|\alpha|^2$, and asymmetry $\Theta$, the probability of data bit-flip errors is calculated after $\Theta$ rounds of measurement. This is done by calculating the mean value of the invariant $J_Z$ on the data cat qubit, defined in \cite{NotesHouches} as $J_Z=J_{+-}+J_{-+}$, and exponentially close to $\text{sign}(\hat x)$. As can be seen in Fig.~\ref{fig:data_bitflips}, even though the data cat qubit has an important amount of leakage due to the fast gates, the bit-flip error probability remains exponentially suppressed with the mean number of photons $|\alpha|^2$. More precisely, increasing the asymmetry leads to an increase in the absolute values of the bit-flip probabilities but does not significantly impact their exponential suppression. This can be qualitatively understood from the fact that the distortion of the state induced by the fast gates is local in phase space, while creating bit-flips requires a transfer of population between left and right half planes in the phase space.

The analysis of the impact of leakage on the measurement errors is more subtle. In this subsection, we investigate it with the same toy model simulation of the circuit in Fig.~\ref{fig:z_correlations_circuit}. Further analysis and simulations of the full QEC logical circuit are provided in the next subsection. Here, both ancilla and data qubits are initialized in the state $|+\rangle=\ket{\cC_\alpha^+}$. In absence of errors, all $\Theta$ measurements would produce the outcome $+1$. However, the phase-flips on the ancilla qubit during the operation of the gate lead to measurement errors (note that in absence of other noise sources the data qubit does not undergo any phase-flip). More precisely, the CNOT gates, while inducing a leakage on data cat qubits, do not change the photon number parity which encodes the logical X operator. This leakage however compromises the functioning of the recurring CNOT gates, leading to further phase-flip errors in the ancilla qubit. This can be seen in Fig.~\ref{fig:control_PhaseFlips} where the probability of ancilla phase flip errors increases after each round of circuit execution. The more subtle effect of the data leakage is however that, this leakage surviving over many measurement rounds, leads to correlated measurement errors. The impact of this correlation can be observed throughout a majority vote as shown in Fig.~\ref{fig:Xmeas_correlations}.

\begin{figure*}[t!] \centering
\sidecaption{fig:z_correlations_circuit}   \raisebox{-\height}{\includegraphics[width=0.20\textwidth,valign=t]{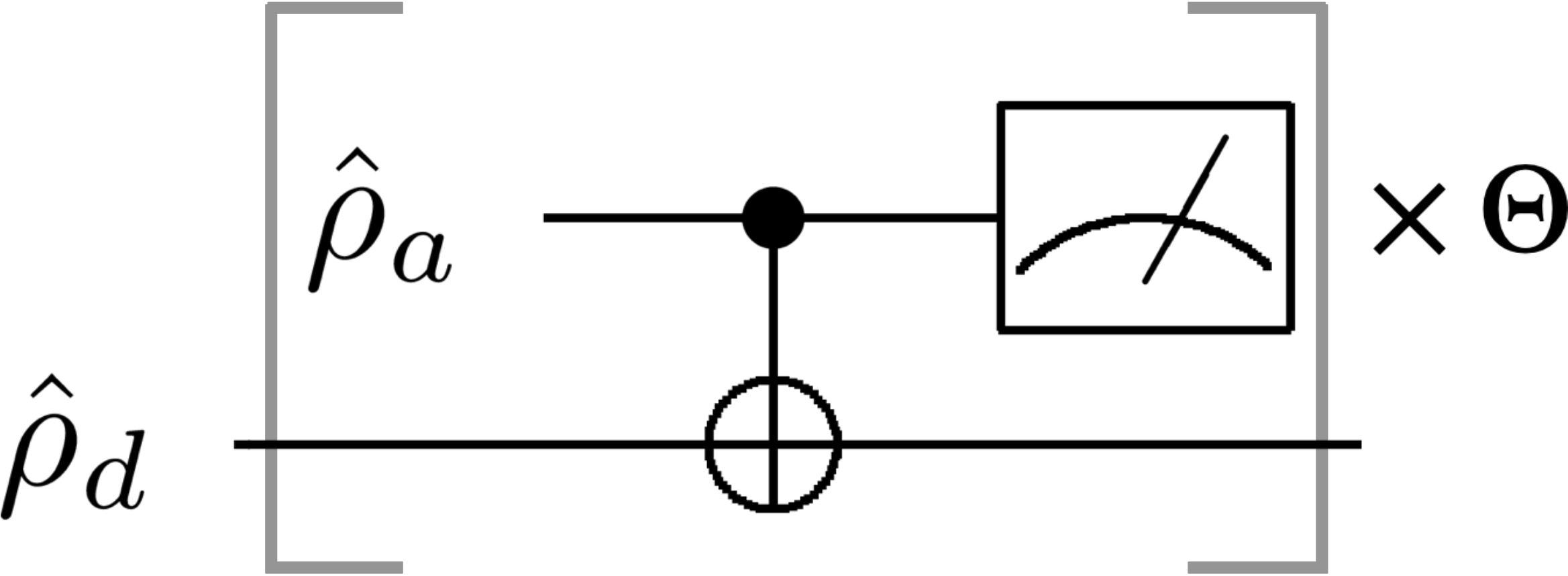} }
\hspace{1ex}
\sidecaption{fig:data_bitflips}   \raisebox{-\height}{\includegraphics[width=0.31\textwidth,valign=t]{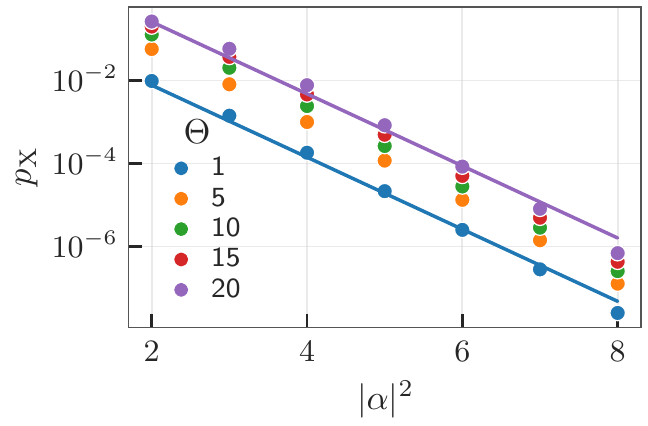} }
\sidecaption{fig:control_PhaseFlips}   \raisebox{-\height}{\includegraphics[width=0.32\textwidth,valign=t]{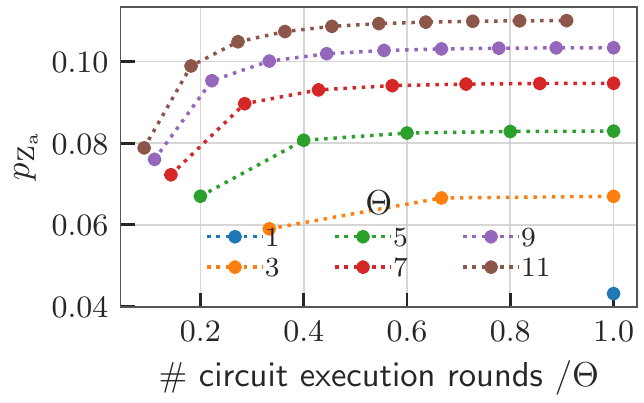} }  \\
\sidecaption{fig:Xmeas_correlations}   \raisebox{-\height}{\includegraphics[width=0.47\textwidth,valign=t]{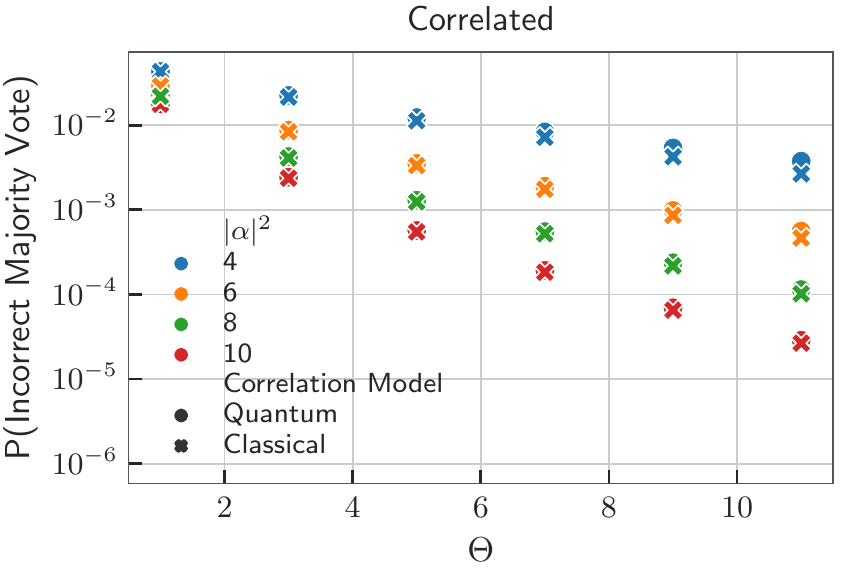} } 
\sidecaption{fig:unXmeas_correlations}   \raisebox{-\height}{\includegraphics[width=0.433\textwidth,valign=t]{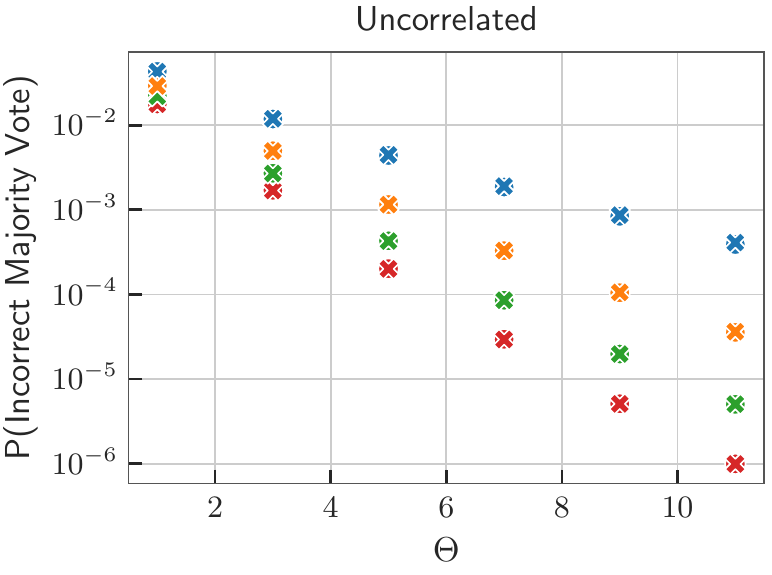} } \\
\caption{(a) Simulated circuit for studying the effect of leakage on bit-flips and phase-flips leading to measurement error correlations. The circuit is repeated $\Theta$ times where $\Theta$ stands for the asymmetry between data and ancilla cat qubit stabilization rates. At each run, the ancilla is re-initialized in the state $\hat \rho_a$ (to be defined later) but the data cat qubit follows its evolution (with leakage out of the cat qubit subspace). (b) Data bit-flip error probability $p_{\mathrm{X}}$ as a function of the mean photon number $|\alpha|^2$  after $\Theta$ execution rounds of circuit in plot (a). In these simulations, we re-initialize the ancilla qubit on the state  $\hat\rho_a= (|+\rangle_C\langle+| + |-\rangle_C\langle-|)/2=(\ket{\cC_\alpha^+}\bra{\cC_\alpha^+}+\ket{\cC_\alpha^-}\bra{\cC_\alpha^-})/2$ at each round. Furthermore, the data one is initialized at the very beginning in the state  $|0\rangle_C=(\ket{\cC_\alpha^+}+\ket{\cC_\alpha^-})/\sqrt{2}$. The two fits  correspond  both to $p_{\mathrm{X}} \propto e^{-2 |\alpha|^2}$ with different pre-factors. (c) Control (ancilla) phase-flip error probability for  $\Theta$ execution rounds of circuit in plot (a) and $|\alpha|^2=4$. As it can be seen, for each fixed $\Theta$, due to the accumulated leakage in data cat-qubits, the probability of these phase-flip errors (leading to measurement errors) increase with the number of execution rounds. Furthermore, the overall probability also increases with $\Theta$ as the  operation becomes faster with respect to data qubits stabilization rate. In these simulations, we re-initialize the ancilla qubit on the state  $\hat\rho_a= |+\rangle_C\langle+| =\ket{\cC_\alpha^+}\bra{\cC_\alpha^+}$ at each round. Furthermore, the data one is initialized at the very beginning in the state  $|+\rangle_C=\ket{\cC_\alpha^+}$. The measurement is performed along the $X$ axis of the ancilla qubit. Ideally each measurement should give the value $+1$, but the phase-flip errors of the ancilla lead to measurement errors whose probability increases with the number of execution rounds and with the asymmetry. (d) This plot studies the leakage-induced  correlation in the measurement errors of plot (c). The plain lines (quantum correlations) in the left plot correspond to the measurement error if we  rely on a majority vote of the $\Theta$ measurement results  of circuit in plot (a). In these simulations, the measurement errors will be correlated as a result of long-lived leakage of data cat qubits.  The right plot corresponds to the same measurement error if the error probabilities are taken from plot (c) but we neglect the potential correlations between them.  We can observe a significant difference between these results pointing towards the importance of these correlations. The dashed line in the left plot correspond to an effective and tractable classical model for the correlations obtained in Section~\ref{ssec:asym_correlation}. We see that this classical model captures quite well the quantum correlations due to the leakage. }
\end{figure*}

The probability of an incorrect majority vote (a majority of `$-1$' measurement outcomes) is plotted in Fig.~\ref{fig:Xmeas_correlations} (left-hand plot), with the label ``Quantum correlations'', as the master equation simulations correspond to a full quantum treatment. These simulations are to be compared to the right-hand plot in the same Figure~\ref{fig:Xmeas_correlations}, where the measurement error correlations are neglected by refocusing the data mode to the cat manifold after each measurement. The higher error probabilities in the left-hand plot reveal the impact of correlation: less information is extracted through each measurement.

The good news however is that even in presence of these correlations, increasing the system asymmetry $\Theta$ leads to a significant improvement of the overall measurement fidelity. Even though larger asymmetry increases the target state leakage, resulting both in individual lower CNOT gate fidelities and correlations between measurement errors, the fact that it is possible to repeat more measurements in the same amount of time $1/\kappa_2^d$ extracts more information. For instance, for a mean photon number of $|\alpha|^2 = 10$, the measurement infidelity in the symmetric case ($\Theta = 1$, and therefore a single measurement) is about $1.8 \times 10^{-2}$; while $\Theta = 11$ (and thus repeating the measurement 11 times) yields an effective measurement infidelity of $2.9\times 10^{-5}$.

 As a conclusion, even though the leakage-induced correlations indeed reduce the global fidelity obtained from an increased measurement repetition rate, using an asymmetric ancilla-data system remains an 
 efficient strategy to obtain a high-fidelity effective measurement. In the next subsection, we go further in this analysis and explain how to capture such correlation effects in full QEC circuit simulations.

\subsection{Tractable model for leakage-induced correlations}
\label{ssec:asym_correlation}

In this subsection, we develop a model to perform circuit-level simulations of a repetition code while including the effect of state leakage. In previous works~\cite{Guillaud_2021,Chamberland2022}, the circuit-level simulations of concatenated `cat qubit + repetition code' were done in two steps: first, an effective Pauli error model was derived for the cat qubits. This was achieved with an analytical model reduction or using a master equation simulation. The goal of this first step is to reduce the description of error channels acting on the full Hilbert space of the harmonic oscillator to a description on the two-dimensional cat qubit manifold. The second step then consists in performing (efficient) sampling of the repetition code logical circuit using these effective error models. 

In the present work, however, we are interested in the regime where the state of data cat qubits are highly deformed and thus cannot be treated as two-dimensional systems. Furthermore, we are specifically interested in investigating the effect of leakage-induced correlations on the logical error probability of the repetition code. Thus, it is crucial to use an enlarged (dimension $>2$) Hilbert space to capture the effect of state leakage. 

The strategy used to perform such simulations is the following. First, we describe the system dynamics in a basis adapted to the cat qubit encoding, the so-called `Shifted Fock Basis' (SFB) introduced in ~\cite{Chamberland2022}. We argue that the quantum coherence created between such basis states can be safely neglected for the purpose of capturing the correlation effects. This assumption is justified both with numerical evidence and by making a model reduction (valid in the regime $\Theta \gg 1$) for which we show explicitly that the dynamics does not create significant coherence between these states. Under this assumption, the errors due to the CNOT process map a pure state to a classical mixture of such basis states. The circuit is then efficiently sampled by generating a random number to select one state of the classical mixture according to the corresponding probability distribution. We now detail these steps.

For a detailed introduction to the SFB, we refer the reader to the Appendix C of~\cite{Chamberland2022}, and we only recall the basics here for self-completeness. The basis is built using two families of `cat-like' states of well-defined photon-number parity based on displaced Fock states
$$
|\phi_{\pm, n}\rangle := \tfrac{1}{\sqrt2}[\mathcal{D}(\alpha)\pm(-1)^n\mathcal{D}(-\alpha)]|n\rangle.
$$
These states are not normalized (but their norm is exponentially close to 1 in the limit $|\alpha|^2 \gg 1$), and the cat qubit subspace is spanned by $|\phi_{\pm, 0}\rangle$. The index $\pm$ refers to the photon-number parity of the associated state, and the index $n$ refers to the excitation number out of the cat qubit subspace. Following the subsystem decomposition idea in~\cite{Pantaleoni_2020}, the basis states may be written as $|\phi_{\pm, n}\rangle = |\pm\rangle \otimes |n\rangle$ where the first state in the tensor product refers to the state of a logical encoded qubit and the second one refers to a gauge mode. Noting that the annihilation operator of the original mode $\hat{a}$ acts as $\hat{a} |\phi_{\pm, n}\rangle = \sqrt{n}|\phi_{\mp, n-1}\rangle + \alpha |\phi_{\mp, n}\rangle$, in the SFB it writes $\hat{a} = Z_a\otimes(\hat{b}_a + \alpha)$ where, $Z_a$ represents the Pauli $Z$ operator on the encoded qubit, and $\hat b_a$ represents the annihilation operator of the virtual gauge mode. In this basis, assuming $\kappa_1^a=\kappa_1^d=0$, the undesired part of the dynamics associated to the CNOT process (obtained by going to an appropriate rotating frame~\cite{Chamberland2022} in~(\ref{eq:ME_init})) can be approximated by 
\begin{equation} \begin{aligned}
\frac{d\rho}{dt}&=-i\frac{\pi}{4T}\left[Z_a\otimes(\hat b_a\hat b_d+\hat b_a^\dag\hat b_d^\dag),\rho\right]\\
&+4\kappa_2^a|\alpha|^2\mathcal{D}[\hat b_a]\rho\\
&+4\kappa_2^d|\alpha|^2\mathcal{D}[Z_a(2\pi t/T)\otimes \hat b_d]\rho,
\end{aligned} \label{eq:ME_CNOT_SFB} \end{equation}
where $Z_a(\theta)=\exp(i\theta Z_a/2)$. In this approximation, detailed in the Appendix D of~\cite{Chamberland2022} (equation (D26)), we consider at most one excitation in the gauge modes and weak couplings are also neglected. More precisely, the above approximate model only makes sense up to the first excitation of the gauge modes $\hat b_a$ and $\hat b_d$. Therefore, the gauge modes $\hat b_a$ and $\hat b_d$ can be replaced by gauge qubits $\hat\sigma^g_{a,-}$ and $\hat\sigma^g_{d,-}$. Here, noting furthermore that $\kappa_2^a\gg \kappa_2^d$, we adiabatically eliminate the ancilla gauge qubit $\hat\sigma^g_{a,-}$, while keeping the data gauge qubit $\hat\sigma^g_{d,-}$. 

\begin{figure*}[t!]\begin{center}
\includegraphics[width=0.95\textwidth]{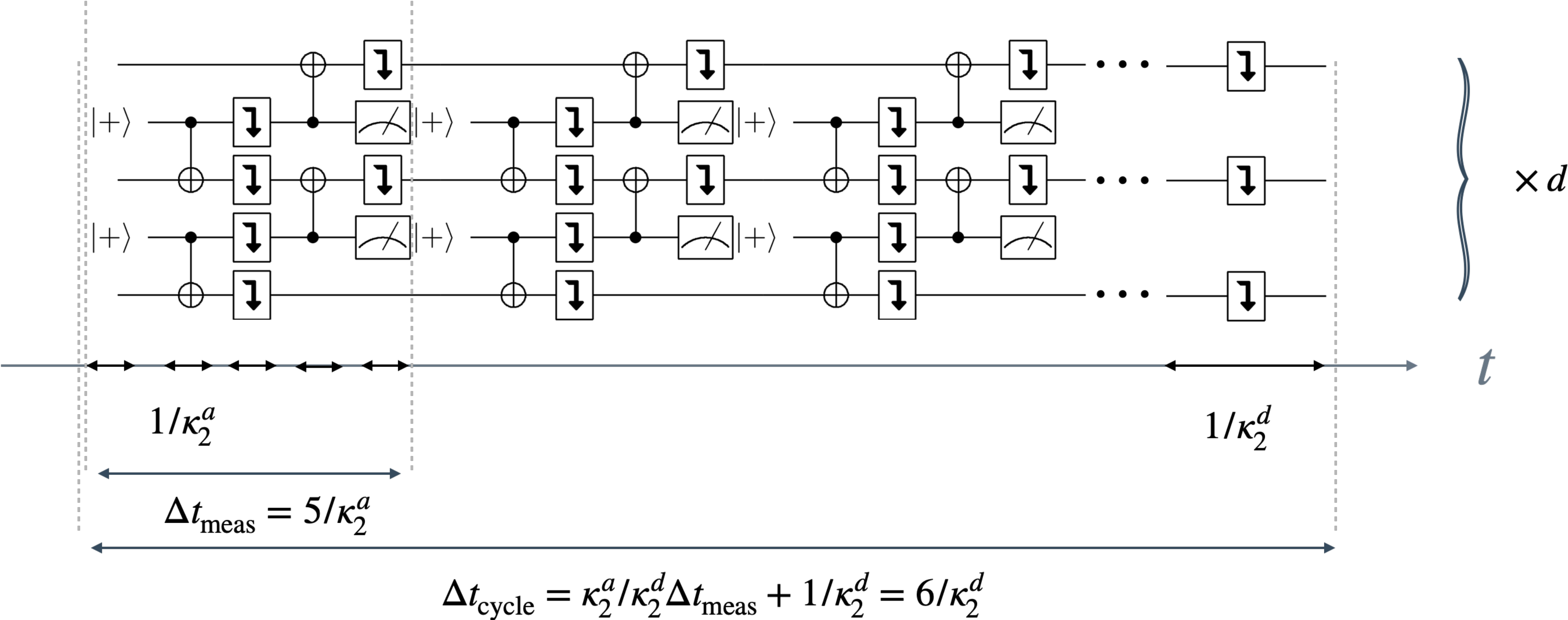}
\end{center} \caption{A quantum error correction cycle for an asymmetric repetition cat qubit (here the code distance is $d=3$). In each cycle, the $d-1$ stabilizers are measured $\Theta$ times, noting that each measurement round takes a duration of $\Delta t_{\text{meas}}=5/\kappa_2^a$. This duration corresponds to an ancilla preparation time step, two rounds of CNOT operations spaced by an ancilla qubit refreshing time step, and finally an ancilla measurement time step. After $\Theta$ rounds of fast measurement cycles, we add a round of data qubit refreshing step of duration $1/\kappa_2^d$, bringing the total duration of the QEC cycle to $\Delta t_{\text{cycle}}=6/\kappa_2^d$. This QEC cycle is repeated $d$ times before going through  a MWPM decoder.}
\label{fig:QEC_circuit_asym} \end{figure*}

This leads to the effective master equation
\begin{equation}\begin{aligned}
\frac{d\rho}{dt}=&4\kappa_2^d|\alpha|^2\mathcal{D}[Z_a(2\pi t/T)\otimes \hat\sigma^g_{d,-}]\rho\\
&+\frac{\pi^2}{16|\alpha|^2\kappa_2^aT^2}\mathcal{D}[Z_a\otimes\hat\sigma^g_{d,+}]\rho.
\end{aligned}\end{equation}\label{eq:Eff_ME_CNOT_SFB}
We note that for an initial state of the form 
$$
\rho_{\text{in}}=\rho_{in,0}^a\otimes\ket{0}_g^d\bra{0}+\rho_{in,1}^a\otimes\ket{1}_g^d\bra{1}
$$
the solution remains of the same form (i.e. diagonal with respect to the data gauge qubit), with $\rho_0$ and $\rho_1$ satisfying
\begin{align}
\frac{d}{dt}\rho^a_0&=r_1Z_a\left(\frac{2\pi t}{T}\right)\rho^a_1Z_a\left(-\frac{2\pi t}{T}\right)-r_2\rho^a_0\notag\\
\frac{d}{dt}\rho^a_1&=r_2 Z_a\rho^a_0 Z_a-r_1\rho^a_1,
\end{align}
with $r_1=4\kappa_2^d|\alpha|^2$ and $r_2=\pi^2/16|\alpha|^2\kappa_2^aT^2$.

The above observation essentially means that the data gauge qubit can be treated as a classical memory bit. More precisely, during each CNOT operation, starting from any state 0 or 1, this classical bit can either stay in its initial state or switch to the other state and this is accompanied by the application of an appropriate partial Kraus map on the ancilla qubit. We denote these Kraus maps as $\bK_{0\rightarrow0}$, $\bK_{0\rightarrow1}$, $\bK_{1\rightarrow0}$, and $\bK_{1\rightarrow1}$, and note that starting from the ancilla qubit state $\rho^a$ and data gauge bit 0, the data gauge bit remains in the state $0$ with probability $\text{tr}\left[\bK_{0\rightarrow0}(\rho)\right]$ and switches from $0$ to $1$ with probability $\text{tr}\left[\bK_{0\rightarrow1}(\rho)\right]$. The validity of these assertions is checked by simulations of Fig.~\ref{fig:Xmeas_correlations}. In these simulations, labeled ``Classical correlations'', we simulate the circuit of Fig.~\ref{fig:z_correlations_circuit}, but this time by treating the data gauge mode as a classical bit. This is done by neglecting the off-diagonal elements of the density matrix in the evolution of the master equation. Such a classical treatment of the data gauge modes will become more clear in the following discussion of the QEC circuit simulations.  

In each error correction round, the ancilla qubits, initialized in $\ket{+}\bra{+}$, undergo two such Kraus maps associated with the adjacent data gauge qubits. These ancilla qubits are finally measured in the $X$ basis. More precisely, the state of ancilla qubit $a$ adjacent to two data gauge bits $d$ and $d'$, undergoes the Kraus maps $\bK_{i'\rightarrow j'}\circ \bK_{i\rightarrow j}$, before being measured in the $X$ basis. Here $i$ and $j$ (resp. $i'$ and $j'$) are initial and final states of the data gauge bit $d$ (resp. $d'$). To perform efficient circuit-level simulations of the repetition code while accounting for state leakage, one therefore only needs to estimate the values $$
p_{i'\rightarrow j',i\rightarrow j}=\bra{-}\bK_{i'\rightarrow j'}\circ \bK_{i\rightarrow j}(\ket{+}\bra{+})\ket{-}.
$$
These values correspond to the probabilities of an erroneous measurement, conditioned to the classical gauge bits $d$ and $d'$ switching respectively from states $i$ and $i'$ to the states $j$ and $j'$. In practice, we evaluate the above probabilities by simulating twice the master equation~\eqref{eq:ME_CNOT_SFB} associated to a CNOT gate, once with data gauge mode $d$ and once with data gauge mode $d'$, where the ancilla gauge mode is initialized in $\ket{0}\bra{0}$ and the ancilla qubit in $\ket{+}\bra{+}$. Furthermore, the data gauge modes $d$ and $d'$ are initialized in $\ket{i}$ and $\ket{i'}$ and we calculate the final population of $\ket{-}_a\otimes\ket{j}_d\otimes\ket{j'}_{d'}$.

\subsection{Overhead estimates \label{ssec:overhead_asym}}

Using the efficient sampling of the repetition code, we perform Monte Carlo simulations of the repetition code to estimate the thresholds and logical phase-flip error rates for increasing system asymmetry. For a symmetric system ($\Theta$=1), the common practice is to repeat $d$ times the stabilizer measurements (followed by a final perfect measurement of the stabilizers to ensure projection over the logical code space) to estimate the logical error probability, where $d$ is the code distance. For an asymmetric system ($\Theta$>1), the temporal correlations between measurement errors result in a decrease of the effective 'time' distance of the code, such that estimating the threshold on a time window of $d$ rounds would be inaccurate. Instead, we replace each of the $d$ stabilizer measurement rounds by a block of $\Theta$ `fast' stabilizer measurements, such that the logical error probability is evaluated over a constant total time, even when the asymmetry is increased. After each block of $\Theta$ `fast' stabilizer measurements, a refreshing time of duration $1/\kappa_2^d$ is inserted on the data cat qubits to remove the leakage. The simulated circuit is summarized in Figure~\ref{fig:QEC_circuit_asym}.

\begin{figure*}[t!] \centering
\sidecaption{subfig:a}   \raisebox{-\height}{\includegraphics[width=0.29\textwidth,valign=t]{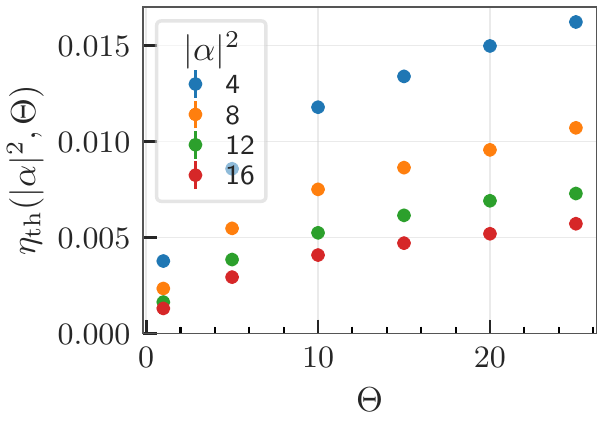}}
\sidecaption{subfig:b}   \raisebox{-\height}{\includegraphics[width=0.29\textwidth,valign=t]{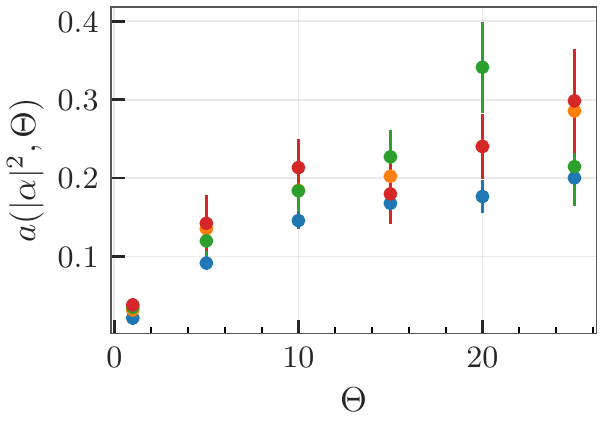}}
\sidecaption{subfig:c}   \raisebox{-\height}{\includegraphics[width=0.29\textwidth,valign=t]{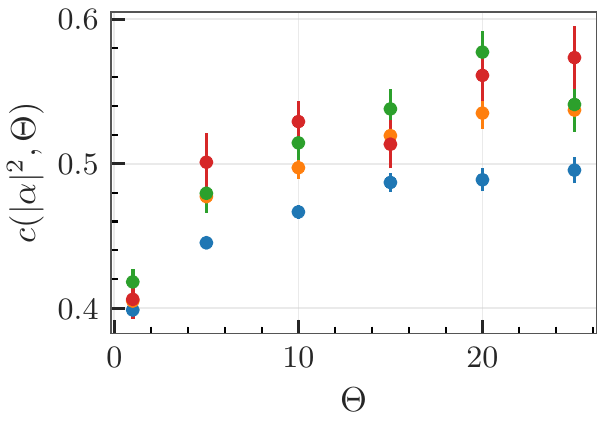}}  \\
\caption{(a) Logical phase-flip error threshold $\eta_{\text{th}}$, (b) prefactor $a$ and (c) scaling coefficient $c$ in the logical phase-flip error probability fit $p_{Z_{\mathrm{L}}} = a d (\eta / \eta_{\mathrm{th}})^{c(d+1)}$ as a function of the asymmetry $\Theta$ for different values of $|\alpha|^2$.
    The error bars are computed as 1.96 standard deviation errors on the parameters (corresponding to 95\% confidence level) from the covariance matrix of the fit and displayed in the three plots\reviews{, see App.\ref{sec:mc}}.
\label{fig:thresholds}}
\end{figure*}

The simulation results are fitted to the empirical formula 
\begin{multline}\label{eq:empirical}
p_{\mathrm{Z_L}}(d, \eta, |\alpha|^2, \Theta) \approx \\
a(|\alpha|^2, \Theta) \left(\frac{\eta}{\eta_{\text{th}}(|\alpha|^2, \Theta)}\right)^{c(|\alpha|^2, \Theta)(d+1)}
\end{multline}
to estimate the phase-flip threshold $\eta_{\text{th}}$, the prefactor $a$, and the scaling coefficient $c$, for different system asymmetries $\Theta$ and different values of the mean photon number $|\alpha|^2$. These estimations are depicted in  Figure~\ref{fig:thresholds}. As expected, the threshold increases with the system asymmetry. Each block of $\Theta$ rounds of stabilizer measurements (replacing a single round in the symmetric case) implements an effective high fidelity stabilizer measurement (as in the case of the X measurement of the previous subsection, see Figure~\ref{fig:Xmeas_correlations}). 

\begin{figure*}[t!] \centering   
\includegraphics[width=\textwidth]{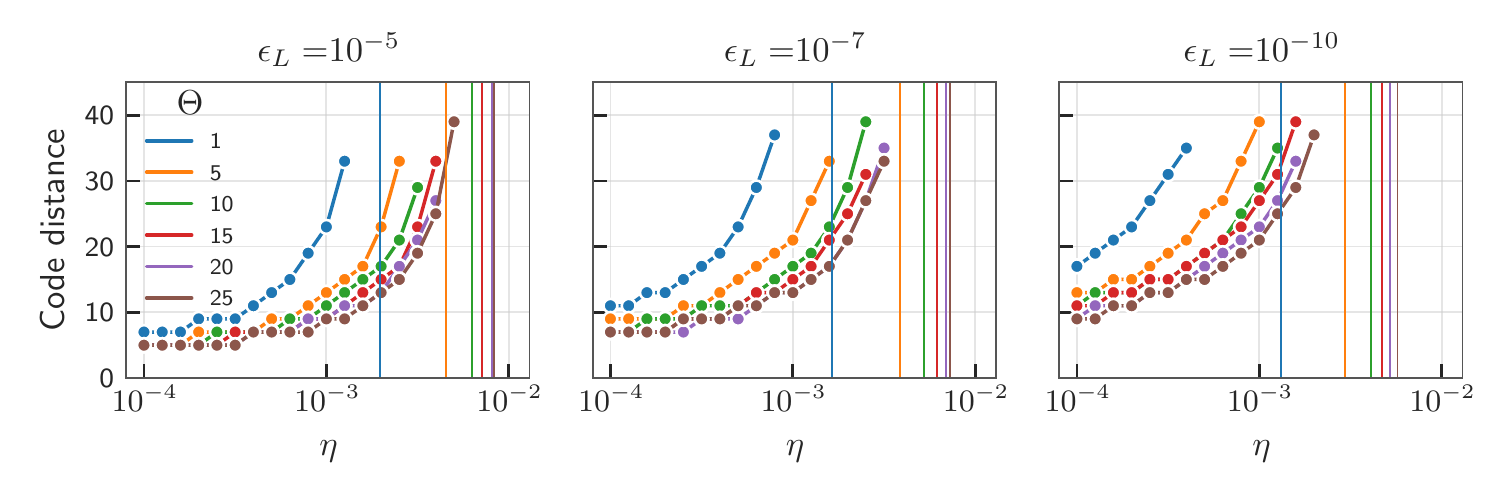}
 \caption{Estimated number of cat qubits per repetition cat qubit used as a quantum memory, versus the physical figure of merit $\eta$. The three plots (left, center and right) correspond to different values of the target logical error probability per QEC cycle ($10^{-5}$, $10^{-7}$ and $10^{-10}$ respectively) and colors correspond  to different values of the asymmetry $\Theta = \kappa_2^a / \kappa_2^d$. For the  points shown in these plots, the required mean number of photons $|\alpha|^2$ to reach the target logical error rates of $10^{-5}$, $10^{-7}$ and $10^{-10}$ are respectively in the ranges $\left[8, 10\right]$, $\left[10, 12\right]$ and $\left[14, 16\right]$. These points are found through a fitting procedure to extrapolate  per-cycle  phase-flip error rates with increasing code distance $d$. For each plot, the minimum  of the phase-flip thresholds $\eta_{\text{th}}(\Theta, |\alpha|^2)$, over $|\alpha|^2$ in the corresponding range, are displayed as vertical asymptotic frontiers. This can be seen as an upper bound for the value of $\eta$ above which one cannot  reach the target error rate.}\label{fig:overhead_asym} \end{figure*}

Finally, we estimate as in Subsection~\ref{ssec:overhead_reducedtime} the overhead required to achieve a per cycle logical error rate of $\epsilon_{\mathrm{L}} = 10^{-5}, 10^{-7}, 10^{-10}$. 
More precisely,  we first use the above fit~\eqref{eq:empirical} for the logical phase-flip  probability, at fixed $(|\alpha|^2, \Theta)$, to extrapolate $p_{\mathrm{Z_L}}$ for larger code distances $d$ and smaller figure of merits $\eta$.
We also estimate the per cycle bit-flip error probability $p_{\mathrm{X_L}}(d, \Theta, |\alpha|^2) \propto d\exp(-2|\alpha|^2)$ (which we find numerically to be dominated by the non-adiabatic bit-flips during the CNOT gates). Finally, for each value of $\eta$, we numerically optimize the code distance $d$ and the average number of photons to achieve $p_{\mathrm{X_L}} + p_{\mathrm{Z_L}} \leq \epsilon_{\mathrm{L}}$. The resulting overheads, for different values of the system asymmetry $\Theta$, are summarized in Figure~\ref{fig:overhead_asym}.

As one could expect from the increase in the repetition code threshold (Figure~\ref{fig:thresholds}), increasing the system asymmetry improves drastically the performance of the repetition code. For instance, for a fixed value of $\eta = 10^{-3}$, a logical error probability of $10^{-10}$ cannot be achieved for a symmetric system, but can be attained with  $d = 25$ data cat qubits of size $|\alpha|^2 \approx 16$ per logical qubit for a system asymmetry of $\Theta = 20$.

\section{Conclusions and further discussions}
\label{sec:Conclusion}

In this work, we proposed and analyzed the acceleration of parity measurement cycle in repetition cat qubits as a means to drastically improve its error correction performance. This acceleration includes two ingredients.

The first ingredient consists in accelerating the CNOT gate, which decreases the fidelity of the gate but perhaps counter-intuitively, improves the overall performance of the code. We explain this improvement by the asymmetric (between control and target) error model of the CNOT gate for cat qubits, and by the fact that the repetition code is more robust to measurement errors than to errors damaging the encoded information. By accelerating the CNOT operations, one however needs to carefully consider the cat qubits state leakage outside the computational subspace. We have analyzed the effects of this state leakage and shown how it can be mitigated by adding appropriate qubit refreshing time steps in the logical circuit.

The second ingredient relies on an asymmetric architecture, where we assume that the typical dissipative rates (both of the stabilization, and of the typical decoherence) of the ancilla cat qubits can be made larger than those of the data cat qubits. To analyze the performance of the repetition code in this regime where data cat qubits suffer from important state leakage, we introduce a new numerical method that allows to efficiently sample the repetition code under a circuit-level noise model, while taking into account the leakage of the cat qubits. The crux of this method was to develop a classical model of correlations that faithfully captures the effect of the leakage-induced correlations in measurement outcomes. We find that this scheme achieves close-to-optimal performance of the repetition code, leading to high values of the phase-flip threshold (\textit{e.g} $\eta_{\text{th}} \approx 1\%$ for $|\alpha|^2 = 8$ photons). 

This proposal is very much inspired by the experimental observation that while the parameter $\eta=\kappa_1/\kappa_2$ is hard to decrease, there is some room for varying the absolute values of these loss rates $\kappa_1$ and $\kappa_2$. One can for instance think of an architecture where the data cat qubits are hosted in extremely high-Q 3D cavity modes, and where the ancilla ones are hosted in lower-Q 2D resonators.  The  performances observed through the Monte Carlo simulations of this paper are encouraging for such a concatenated and asymmetric architecture.

Throughout this work, we have analyzed exclusively the logical performance of a quantum memory. One may legitimately wonder if the same conclusions still apply to the case of logical gate implementations for repetition cat qubits. In this architecture~\cite{Guillaud_2019,Guillaud_2021,Chamberland2022} two types of gate implementations can be distinguished: the transversal ones such as the CNOT gate and the non-transversal ones such as the Toffoli. For the transversal implementations, we expect a similar improvement in the logical performance with fast noisy gates. More precisely, while there is no interest in accelerating the CNOT operations between the data qubits in two code blocks, the parity-check CNOTs in each block can still benefit from the same acceleration. The CNOT gate errors between data qubits merely act as the input errors of a memory. The code being  resilient to these input errors, the logical fidelity of the transversal gate is mainly limited by the performance of the error correction circuit.  The analysis for the non-transversal implementations is less straight-forward and requires further investigation. However, we believe that with some modifications, these implementations can also benefit from overhead reduction using fast noisy parity-checks.  

One possible direction for extending this work is to consider  biased noise tailored codes that have some bit-flip error correction capability~\cite{Chamberland2022, Darmawan2021}. In this case, one needs to keep in mind that the measurement of $Z$-stabilizers would require CZ or CNOT gates with data qubits as control ones. The data qubits are thus necessarily affected by the non-adiabatic errors and as such one cannot rely on fast low-fidelity gates for $Z$-stabilizer measurements. It should however be possible to rely on two time scales, one  fast  for $X$-stabilizer measurements as they need to compete with high-rate $Z$ errors, and one slow for $Z$-stabilizers competing with rare $X$ errors.
\section{Acknowledgments}

We thank Alain Sarlette for insightful discussions. We acknowledge funding from the Plan France 2030 through the project ANR-22-PETQ-0006. 

\bibliographystyle{quantum}
\bibliography{main}

\begin{thebibliography}{10}

\bibitem{Joshi2021}
Atharv Joshi, Kyungjoo Noh, and Yvonne~Y Gao.
\newblock ``Quantum information processing with bosonic qubits in circuit {QED}''.
\newblock \href{https://dx.doi.org/10.1088/2058-9565/abe989}{Quantum Science and Technology {\bf 6}, 033001}~(2021).

\bibitem{Cai2021}
Weizhou Cai, Yuwei Ma, Weiting Wang, Chang-Ling Zou, and Luyan Sun.
\newblock ``Bosonic quantum error correction codes in superconducting quantum circuits''.
\newblock \href{https://dx.doi.org/10.1016/j.fmre.2020.12.006}{Fundamental Research {\bf 1}, 50--67}~(2021).

\bibitem{Mirrahimi2014}
Mazyar Mirrahimi, Zaki Leghtas, Victor~V Albert, Steven Touzard, Robert~J Schoelkopf, Liang Jiang, and Michel~H Devoret.
\newblock ``Dynamically protected cat-qubits: a new paradigm for universal quantum computation''.
\newblock \href{https://dx.doi.org/10.1088/1367-2630/16/4/045014}{New Journal of Physics {\bf 16}, 045014}~(2014).

\bibitem{Guillaud_2019}
Jérémie Guillaud and Mazyar Mirrahimi.
\newblock ``Repetition cat qubits for fault-tolerant quantum computation''.
\newblock \href{https://dx.doi.org/10.1103/physrevx.9.041053}{Physical Review X{\bf 9}}~(2019).

\bibitem{Chamberland2022}
Christopher Chamberland, Kyungjoo Noh, Patricio Arrangoiz-Arriola, Earl~T. Campbell, Connor~T. Hann, Joseph Iverson, Harald Putterman, Thomas~C. Bohdanowicz, Steven~T. Flammia, Andrew Keller, Gil Refael, John Preskill, Liang Jiang, Amir~H. Safavi-Naeini, Oskar Painter, and Fernando~G.S.L. Brandão.
\newblock ``Building a fault-tolerant quantum computer using concatenated cat codes''.
\newblock \href{https://dx.doi.org/10.1103/prxquantum.3.010329}{PRX Quantum{\bf 3}}~(2022).

\bibitem{Lescanne2020}
Raphaël Lescanne, Marius Villiers, Théau Peronnin, Alain Sarlette, Matthieu Delbecq, Benjamin Huard, Takis Kontos, Mazyar Mirrahimi, and Zaki Leghtas.
\newblock ``Exponential suppression of bit-flips in a qubit encoded in an oscillator''.
\newblock \href{https://dx.doi.org/10.1038/s41567-020-0824-x}{Nature Physics {\bf 16}, 509–513}~(2020).

\bibitem{Berdou2022}
C.~Berdou, A.~Murani, U.~Réglade, W.C. Smith, M.~Villiers, J.~Palomo, M.~Rosticher, A.~Denis, P.~Morfin, M.~Delbecq, T.~Kontos, N.~Pankratova, F.~Rautschke, T.~Peronnin, L.-A. Sellem, P.~Rouchon, A.~Sarlette, M.~Mirrahimi, P.~Campagne-Ibarcq, S.~Jezouin, R.~Lescanne, and Z.~Leghtas.
\newblock ``One hundred second bit-flip time in a two-photon dissipative oscillator''.
\newblock \href{https://dx.doi.org/10.1103/prxquantum.4.020350}{PRX Quantum{\bf 4}}~(2023).

\bibitem{Lutterbach1997}
L.~G. Lutterbach and L.~Davidovich.
\newblock ``Method for direct measurement of the wigner function in cavity qed and ion traps''.
\newblock \href{https://dx.doi.org/10.1103/PhysRevLett.78.2547}{Phys. Rev. Lett. {\bf 78}, 2547--2550}~(1997).

\bibitem{Bertet2002}
P.~Bertet, A.~Auffeves, P.~Maioli, S.~Osnaghi, T.~Meunier, M.~Brune, J.~M. Raimond, and S.~Haroche.
\newblock ``Direct measurement of the wigner function of a one-photon fock state in a cavity''.
\newblock \href{https://dx.doi.org/10.1103/PhysRevLett.89.200402}{Phys. Rev. Lett. {\bf 89}, 200402}~(2002).

\bibitem{Sun2014}
L.~Sun, A.~Petrenko, Z.~Leghtas, B.~Vlastakis, G.~Kirchmair, K.~M. Sliwa, A.~Narla, M.~Hatridge, S.~Shankar, J.~Blumoff, L.~Frunzio, M.~Mirrahimi, M.~H. Devoret, and R.~J. Schoelkopf.
\newblock ``Tracking photon jumps with repeated quantum non-demolition parity measurements''.
\newblock \href{https://dx.doi.org/10.1038/nature13436}{Nature {\bf 511}, 444--+}~(2014).

\bibitem{Fowler2012}
Austin~G. Fowler, Matteo Mariantoni, John~M. Martinis, and Andrew~N. Cleland.
\newblock ``Surface codes: Towards practical large-scale quantum computation''.
\newblock \href{https://dx.doi.org/10.1103/PhysRevA.86.032324}{Phys. Rev. A {\bf 86}, 032324}~(2012).

\bibitem{Puri2020}
Shruti Puri, Lucas St-Jean, Jonathan~A. Gross, Alexander Grimm, Nicholas~E. Frattini, Pavithran~S. Iyer, Anirudh Krishna, Steven Touzard, Liang Jiang, Alexandre Blais, Steven~T. Flammia, and S.~M. Girvin.
\newblock ``Bias-preserving gates with stabilized cat qubits''.
\newblock \href{https://dx.doi.org/10.1126/sciadv.aay5901}{Science Advances{\bf 6}}~(2020).

\bibitem{Xu2021}
Qian Xu, Joseph~K. Iverson, Fernando G. S.~L. Brand\~ao, and Liang Jiang.
\newblock ``Engineering fast bias-preserving gates on stabilized cat qubits''.
\newblock \href{https://dx.doi.org/10.1103/PhysRevResearch.4.013082}{Phys. Rev. Res. {\bf 4}, 013082}~(2022).

\bibitem{Putterman2022}
Harald Putterman, Joseph Iverson, Qian Xu, Liang Jiang, Oskar Painter, Fernando~G.{\hspace{0.167em}}S.{\hspace{0.167em}}L. Brand{\~{a}}o, and Kyungjoo Noh.
\newblock ``Stabilizing a bosonic qubit using colored dissipation''.
\newblock \href{https://dx.doi.org/10.1103/physrevlett.128.110502}{Physical Review Letters{\bf 128}}~(2022).

\bibitem{Gautier2022}
Ronan Gautier, Alain Sarlette, and Mazyar Mirrahimi.
\newblock ``Combined dissipative and hamiltonian confinement of cat qubits''.
\newblock \href{https://dx.doi.org/10.1103/PRXQuantum.3.020339}{PRX Quantum {\bf 3}, 020339}~(2022).

\bibitem{Ruiz2023}
Diego Ruiz, Ronan Gautier, Jérémie Guillaud, and Mazyar Mirrahimi.
\newblock ``Two-photon driven kerr quantum oscillator with multiple spectral degeneracies''.
\newblock \href{https://dx.doi.org/10.1103/physreva.107.042407}{Physical Review A{\bf 107}}~(2023).

\bibitem{Xu2022Squeezed}
Qian Xu, Guo Zheng, Yu-Xin Wang, Peter Zoller, Aashish~A. Clerk, and Liang Jiang.
\newblock ``Autonomous quantum error correction and fault-tolerant quantum computation with squeezed cat qubits''.
\newblock \href{https://dx.doi.org/10.1038/s41534-023-00746-0}{npj Quantum Information {\bf 9}, 78}~(2023).

\bibitem{Guillaud_2021}
Jérémie Guillaud and Mazyar Mirrahimi.
\newblock ``Error rates and resource overheads of repetition cat qubits''.
\newblock \href{https://dx.doi.org/10.1103/physreva.103.042413}{Physical Review A{\bf 103}}~(2021).

\bibitem{Fowler_PRL_2012}
Austin~G. Fowler, Adam~C. Whiteside, and Lloyd C.~L. Hollenberg.
\newblock ``Towards practical classical processing for the surface code''.
\newblock \href{https://dx.doi.org/10.1103/PhysRevLett.108.180501}{Phys. Rev. Lett. {\bf 108}, 180501}~(2012).

\bibitem{pymatching}
Oscar Higgott.
\newblock ``Pymatching: A python package for decoding quantum codes with minimum-weight perfect matching''.
\newblock \href{https://dx.doi.org/10.1145/3505637}{ACM Transactions on Quantum Computing}~(2021).

\bibitem{Aliferis2007}
P.~Aliferis and B.M. Terhal.
\newblock ``Fault-tolerant quantum computation for local leakage faults''.
\newblock \href{https://dx.doi.org/10.26421/qic7.1-2-9}{Quantum Information and Computation {\bf 7}, 139--156}~(2007).

\bibitem{Battistel2021}
F.~Battistel, B.M. Varbanov, and B.M. Terhal.
\newblock ``Hardware-efficient leakage-reduction scheme for quantum error correction with superconducting transmon qubits''.
\newblock \href{https://dx.doi.org/10.1103/prxquantum.2.030314}{{PRX} Quantum{\bf 2}}~(2021).

\bibitem{McEwen2021}
M.~McEwen et~al.
\newblock ``Removing leakage-induced correlated errors in superconducting quantum error correction''.
\newblock \href{https://dx.doi.org/10.1038/s41467-021-21982-y}{Nature Communications{\bf 12}}~(2021).

\bibitem{Chen2021}
Zijun Chen et~al.
\newblock ``{Exponential suppression of bit or phase errors with cyclic error correction}''.
\newblock \href{https://dx.doi.org/10.1038/s41586-021-03588-y}{Nature {\bf 595}, 383--387}~(2021).

\bibitem{Dennis2002}
Eric Dennis, Alexei Kitaev, Andrew Landahl, and John Preskill.
\newblock ``Topological quantum memory''.
\newblock \href{https://dx.doi.org/10.1063/1.1499754}{Journal of Mathematical Physics {\bf 43}, 4452--4505}~(2002).

\bibitem{NotesHouches}
Jérémie Guillaud, Joachim Cohen, and Mazyar Mirrahimi.
\newblock ``Quantum computation with cat qubits''.
\newblock \href{https://dx.doi.org/10.21468/scipostphyslectnotes.72}{SciPost Physics Lecture Notes}~(2023).

\bibitem{Pantaleoni_2020}
Giacomo Pantaleoni, Ben~Q. Baragiola, and Nicolas~C. Menicucci.
\newblock ``Modular bosonic subsystem codes''.
\newblock \href{https://dx.doi.org/10.1103/physrevlett.125.040501}{Physical Review Letters{\bf 125}}~(2020).

\bibitem{Darmawan2021}
Andrew~S. Darmawan, Benjamin~J. Brown, Arne~L. Grimsmo, David~K. Tuckett, and Shruti Puri.
\newblock ``Practical quantum error correction with the {XZZX} code and kerr-cat qubits''.
\newblock \href{https://dx.doi.org/10.1103/prxquantum.2.030345}{{PRX} Quantum{\bf 2}}~(2021).

\bibitem{Aaronson2004}
Scott Aaronson and Daniel Gottesman.
\newblock ``Improved simulation of stabilizer circuits''.
\newblock \href{https://dx.doi.org/10.1103/PhysRevA.70.052328}{Phys. Rev. A {\bf 70}, 052328}~(2004).

\bibitem{Edmonds1965}
Jack Edmonds.
\newblock ``Paths, trees, and flowers''.
\newblock \href{https://dx.doi.org/10.4153/CJM-1965-045-4}{Canadian Journal of Mathematics {\bf 17}, 449--467}~(1965).

\bibitem{Kolmogorov2009}
Vladimir Kolmogorov.
\newblock ``Blossom v: a new implementation of a minimum cost perfect matching algorithm''.
\newblock \href{https://dx.doi.org/10.1007/s12532-009-0002-8}{Mathematical Programming Computation {\bf 1}, 43--67}~(2009).

\end{thebibliography}

\onecolumngrid
\appendix

\section{\reviews{QEC circuit sampling and threshold estimation}}
\label{sec:mc}
\reviews{The results of Figures~\ref{fig:Pheno_threshold_curve},~\ref{fig:fits}, \ref{fig:thresholds} are obtained by Monte Carlo simulations of the associated QEC circuits, given the error model of the gates, the code distance $d$, and the noise parameter. 
}

\reviews{
In general, simulating {stabilizer} QEC circuits is efficient because they rely on Clifford gates, hence one can use the CHP algorithm~\cite{Aaronson2004} \reviews{for sampling from such circuits}.
For the repetition code circuit shown in Fig.~\ref{fig:repetition_cat_code}, the simulations are even simpler as this code is a classical one.
}

\reviews{In all above simulations, the system is initialized in the $+1$-eigenstate of all stabilizers, imperfect stabilizer measurements are repeated $d$ times where $d$ stands for the code distance, and are followed by a last  perfect round of stabilizer measurements, projecting perfectly the code on an error syndrome subspace. This repetition of stabilizer measurements allow us to take into account the measurement error in the decoding procedure. }

\reviews{
We simulate the noisy circuit $N$ times and after each execution, we process the results of the syndrome measurements using a minimum weight perfect matching (MWPM) decoder~\cite{Fowler2012}.
The decoder relies on the precomputed detection graph where the nodes correspond to the locations of detection events, i.e. measurements in the circuit, and the weights are obtained from the errors models of the gates.
Then, calling Dijkstra's shortest path algorithm for each pair of detection events gives the complete subgraph to feed to the matching algorithm, a Blossom V implementation of MWPM algorithm~\cite{Edmonds1965, Kolmogorov2009, pymatching}.
Finally, monitoring all the pairs of matched nodes that account for data qubit errors provides the necessary correction \reviews{step} to remove these errors. \reviews{After this correction step, either the logical state has not changed $\ket{\psi_{\text{out}}}_L = \ket{\psi_{\text{in}}}_L $, in which case the error correction was successful, or a $Z_L$ error has occurred, $\ket{\psi_{\text{out}}}_L = Z_L \ket{\psi_{\text{in}}}_L $.}
}

\reviews{
We thus estimate the logical error probability of the circuit $p_L(d, p)$ as:
$p_L \approx N_{\text{fail}} / N$, where $N_{\text{fail}}$ is the number of samples ending with a logical $Z_L$ error. 
For this study, simulated circuits with circuit-level (respectively phenomenological) error model are run until either $N_{\text{fail}}=500$ (resp. $N_{\text{fail}}=10^3$) logical failures are observed or $N=10^6$ runs were performed (resp. $N=10^7$).
}

\begin{figure}
  \centering
    \begin{overpic}[
      width=0.5\textwidth,
      ]{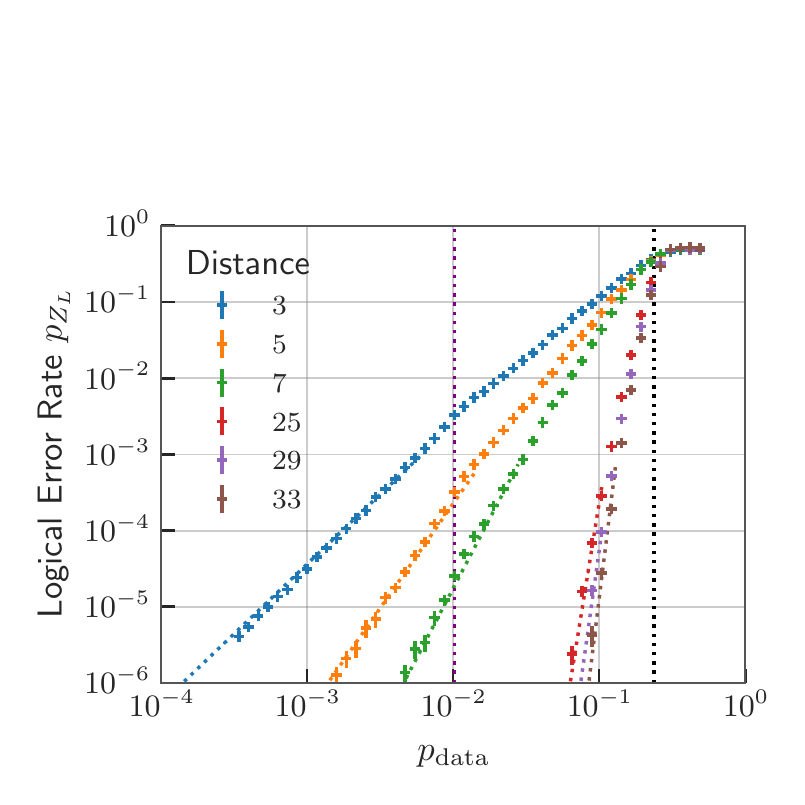}
    \end{overpic}
    \caption{
      \reviews{
      Probability that the error correction based on the stabilizer measurement circuit displayed in
 the inset of Fig.~\ref{fig:Pheno_threshold_curve} using a phenomenological error model with a fixed measurement error of $1\%$ induces a logical $Z_L$
 error on the repetition cat qubit after the correction is performed. The
 dotted lines correspond to the asymptotic regime and fit the empirical
 scaling formula $p_{\mathrm{Z_{L}}}=ad\left(\frac{p_{\mathrm{data}}}{p_{\mathrm{data, th}}}\right)^{c(d+1)}$.
  The purple vertical dotted line is the value of the measurement error $p_{\mathrm{meas}}$ and the black dotted line correspond to the fitted value of the threshold.
      }
    \label{fig:Pheno_threshold_curve_pZL}
    }
\end{figure}

\begin{figure*}[t!] \centering
\sidecaption{subfig:pth_coeff}   \raisebox{-\height}{\includegraphics[width=0.29\textwidth,valign=t]{figures/Pheno_threshold_curve_per_round_pth.pdf}}
\sidecaption{subfig:a_coeff}   \raisebox{-\height}{\includegraphics[width=0.29\textwidth,valign=t]{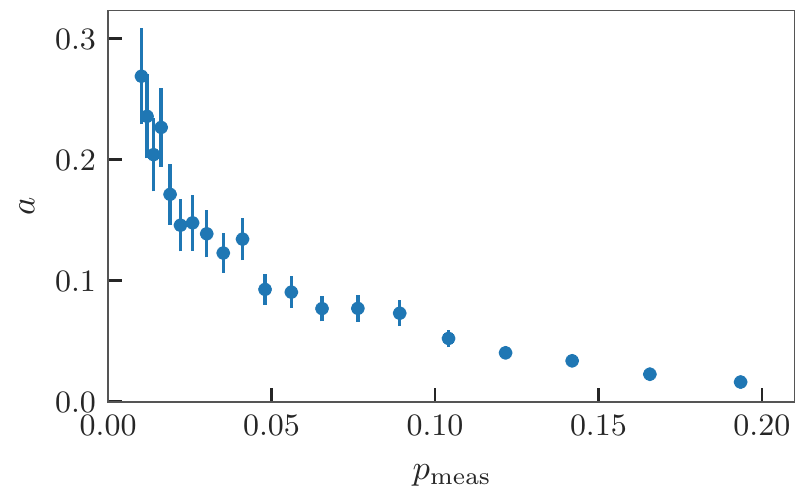}}
\sidecaption{subfig:c_coeff}   \raisebox{-\height}{\includegraphics[width=0.29\textwidth,valign=t]{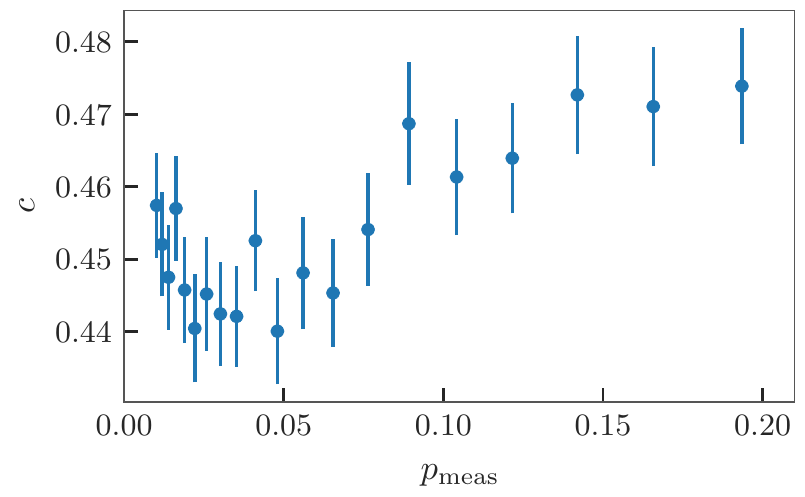}}  \\
\caption{
  \reviews{(a) Logical phase-flip error threshold $p_{\text{data, th}}$ plotted in Figure~\ref{fig:Pheno_threshold_curve}, (b) prefactor $a$ and (c) scaling coefficient $c$ in the logical phase-flip error probability fit $p_{Z_{\mathrm{L}}} = a d (p_{\mathrm{data}} / p_{\mathrm{data, th}})^{c(d+1)}$ as a function of the measurement error $p_{\text{meas}}$.
    The error bars are computed as 1.96 standard deviation errors on the parameters (corresponding to 95\% confidence level) from the covariance matrix of the fit and displayed in the three plots.
    We find a scaling coefficient almost constant and close to its expected value of $1/2$, corresponding to the minimal weight of physical errors leading to a logical one.
  }
 \label{fig:data_per_round}
  }
\end{figure*}

\reviews{In the case of Figure~\ref{fig:Pheno_threshold_curve}, the noise parameter is given by $p_{\mathrm{data}}$. For the sake of completeness, we provide the result of such simulations in the case of $p_{\textrm{meas}}=1\%$ in Figure~\ref{fig:Pheno_threshold_curve_pZL}. In order to obtain the error threshold values plotted in Figure~\ref{fig:Pheno_threshold_curve}, we have fitted the the logical error probabilities in Figure~\ref{fig:Pheno_threshold_curve_pZL} to the ansatz
$p_{Z_{\mathrm{L}}} = a d (p_{\mathrm{data}} / p_{\mathrm{data, th}})^{c(d+1)}.$
As can be seen in Figure~\ref{fig:Pheno_threshold_curve_pZL}, this ansatz fits very well the asymptotic behaviour of the  estimated logical error probabilities. Furthermore,  the values of fit parameters $a$, $c$, and $p_{\mathrm{data, th}}$  are shown in Fig.~\ref{fig:data_per_round} as a function of $p_{\mathrm{meas}}$.
}

\reviews{In the case of Figure~\ref{fig:fits}, the QEC circuits are sampled using a circuit-level error model for the case of cat qubits and bias-preserving operations and the results are fitted to a similar ansatz $p_{Z_L}=ad(\eta/\eta_{\textrm{th}})^{cd}$ as discussed in the caption of the Figure. This formula is then used to extrapolate the results of Figure~\ref{fig:overhead}.}

\reviews{Finally, in the case of Figure~\ref{fig:thresholds} similar simulations are performed for the QEC circuit with the asymmetric ancilla-data error model and including the correlations induced by the data leakage. The obtained logical error probabilities are fitted to the ansatz~\eqref{eq:empirical} and this formula is used to extrapolate the overhead estimates in Figure~\ref{fig:overhead_asym}.}

\end{document}